\documentclass[aps, twocolumn, prl, floatfix, superscriptaddress]{revtex4-2}

\usepackage{amsfonts}
\usepackage{amssymb,amstext,amsthm,amsmath}
\usepackage{mathtools}
\usepackage{graphicx}
\usepackage{enumerate}
\usepackage{physics}
\usepackage{enumitem}

\usepackage[colorlinks=true, allcolors=blue]{hyperref}

\usepackage{multibib} 

\providecommand{\xpct}[1]{\ensuremath{\langle#1\rangle}}

\begin{document}

\title{Quantum Origin of Limit Cycles, Fixed Points, and Critical Slowing Down}

\newcommand{\mpipks}{Max-Planck-Institut für Physik Komplexer Systeme, D-01187 Dresden, Germany}

\author{Shovan Dutta}
\affiliation{\mpipks}
\affiliation{Raman Research Institute, Bangalore 560080, India}

\author{Shu Zhang}
\affiliation{\mpipks}

\author{Masudul Haque}
\affiliation{Institut f\"ur Theoretische Physik, Technische Universit\"at Dresden, 01062 Dresden, Germany}
\affiliation{\mpipks}

\begin{abstract}
Among the most iconic features of classical dissipative dynamics are persistent limit-cycle
oscillations and critical slowing down at the onset of such oscillations, where the
system relaxes purely algebraically in time. On the other hand, quantum systems subject to
generic Markovian dissipation decohere exponentially in time, approaching a unique steady state. Here we show how coherent limit-cycle oscillations and algebraic decay can emerge in a quantum system governed by a Markovian master equation as one approaches the classical limit, illustrating general mechanisms using a single-spin model and a two-site lossy Bose-Hubbard model.  In particular, we demonstrate that the fingerprint of a limit cycle is a slow-decaying branch with vanishing decoherence rates in the Liouville spectrum, while a power-law decay is realized by a spectral collapse at the bifurcation point. We also show how these are distinct from the case of a classical fixed point, for which the quantum spectrum is gapped and can be generated from the linearized classical dynamics.

\end{abstract}

\maketitle

\emph{Introduction.}|The question of how classical behavior appears as an appropriate limit of quantum dynamics has intrigued physicists since the dawn of quantum mechanics~\cite{Ehrenfest1927, Hepp_CommMathPhys1974, Yaffe_RMP1982_largeNlimits, Reichl_Lin_FoundPhys1987, book_Gutzwiller1990chaos,
Zurek_PhysicsToday1991, Zurek_RMP2003_decoherence, book_Schlosshauer_2007_decoherence, Klein_AJP2012_classicallimit}.  For \emph{Hamiltonian}  classical systems, there is a distinction
between chaotic and nonchaotic dynamics: a long line of work
has explored the spectral manifestations of this distinction
in corresponding quantum systems 
~\cite{Bohigas_PRL1984, Robb_Reichl_PRE1998_twospin, Emerson_Ballentine_PRA2001_twospin, Mueller_Braun_Haake_Altland_PRL2004,  Mueller_Altland_Braun_Haake_NJP2009, Hirsch_PRA2014_Dicke_comparative2, Nakerst_Haque_PRE2023, Benet_Borgonovi_Izrailev_Santos_PRB2023, Novotny_Stransky_PRE2023_otoc, Santos_Hirsch_PRL2019_LyapunovOTOCDickemodel, Rautenberg_Gaertner_PRA2020,  Pappalardi_Polkovnikov_Silva_SciPost2020_SherringtonKirkpatrick, Pappalardi_PRA2020_semiclassicalsystems_kickedtop_dickemodel, book_Reichl_TransitionToChaos}.
Non-Hamiltonian (dissipative) classical systems support far more intriguing 
nonlinear phenomena, such as limit cycles, bifurcations, period doubling transitions to chaos, and strange attractors \cite{book_Strogatz_2018, book_Hilborn_2000_chaos}.  However, a  detailed  understanding based on the  quantum spectrum, at the level available for Hamiltonian systems, is currently missing in the dissipative case.  
The present work is a step toward formulating and addressing these questions.

A Markovian quantum dissipative system is described via a master equation $\dot{\rho} = \mathcal{L}\rho$ for the density matrix $\rho$, where the Liouvillian  $\mathcal{L}$ is constrained to have the Lindblad or Gorini–Kossakowski–Sudarshan–Lindblad (GKSL) form \cite{Lindblad1976, GKSL_1976, book_BreuerPetruccione_2002}.
Lindblad dynamics generally relax exponentially in time to a unique steady state, except in the case of special symmetries \cite{Evans1977, Buca_Prosen_NJP2012, Albert_Jiang_PRA2014_Lindbladsymmetries, Nonabelian_Zhang2020}.  The behaviors at a classical limit cycle (persistent oscillations) and at bifurcation points (algebraic decay) appear to contradict this quantum description. Here, we show how such properties emerge from spectral features of $\mathcal{L}$ as the classical limit is approached.  We illustrate our results primarily through a dissipative nonlinear spin model.  To demonstrate that our findings are generic, we also illustrate some of these results using a driven-dissipative two-site Bose-Hubbard model \cite{giraldo2020driven, giraldo2021chaotic, giraldo2022semiclassical}.

\begin{figure}[tbp]
    \centering
    \includegraphics[width=1\columnwidth]{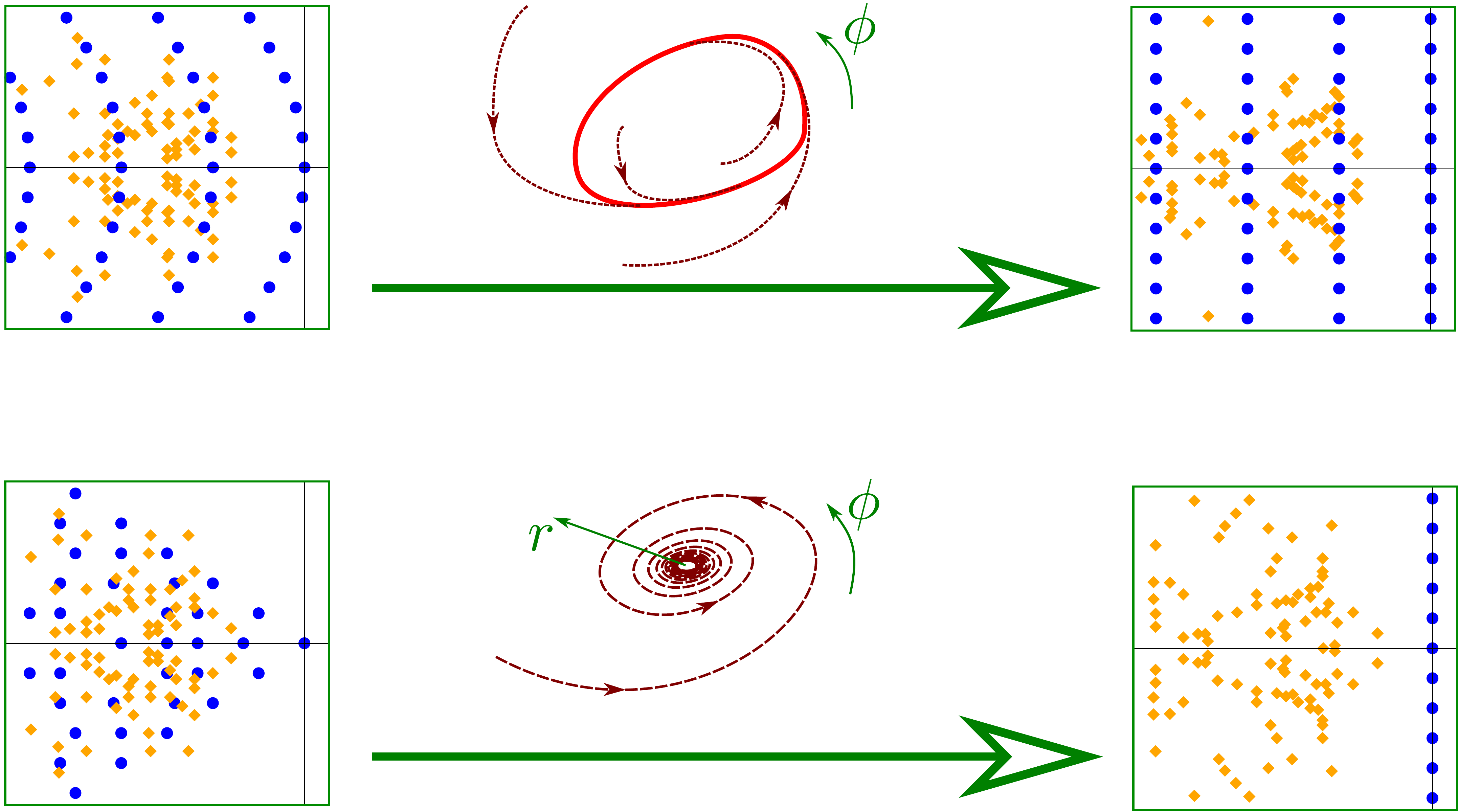}
    \caption{\label{fig:introfig}Schematics illustrating the fate of Liouvillian spectra (left panels) in approaching the classical limit (right panels).  For a limit cycle (top panels) the signature is a branch of equally spaced imaginary eigenvalues yielding coherent oscillations, plus gapped parallel branches, describing the approach to the limit cycle. The branches are parabolic in the quantum case, which causes dephasing at long times. At a Hopf bifurcation point (bottom panels) the spectrum collapses onto the imaginary axis, with macroscopic degeneracies at each harmonic, leading to algebraic decay.  Additional eigenvalues, e.g., associated with other attractors in phase space, may exist, as indicated by the orange diamonds. 
    }
\end{figure}

\emph{Highlights of main results.}|Figure\ \ref{fig:introfig} shows schematics of the Liouvillian spectrum of a system approaching the classical limit that features a limit cycle (top panels) or its onset at a Hopf bifurcation (bottom panels).  We focus on the low-lying (i.e., slow-decaying) part of the spectrum close to the imaginary axis.  The spectral signature of a limit cycle is an approximately parabolic branch of eigenvalues including the steady state, which collapses onto the imaginary axis in the classical limit, forming an equally spaced linear array. The eigenstates of this branch all have the same radial structure in classical  phase space (concentrated around the limit cycle), but each has a different angular structure, $\sim{e}^{{\rm i} l\phi}$ with $l\in\mathbb{Z}$.  The superposition of an infinite number of such states can yield a localized packet in phase space orbiting the limit cycle, recovering the classical late-time dynamics.  Additionally, there are parallel branches that have weights at increasing radial distances with proportional decay rates.  

At a Hopf bifurcation the classical dynamics features critical slowing down, relaxing algebraically in time.  In contrast, in quantum Lindblad dynamics the decay to the steady state is generically exponential, governed by the real Liouvillian gap. This conflict is resolved through an  infinite number of eigenvalues with the same angular but different radial structures collapsing on the imaginary axis in the classical limit.  This allows the radial approach to have a power-law form.  

\begin{figure}
    \includegraphics[width=1\columnwidth]{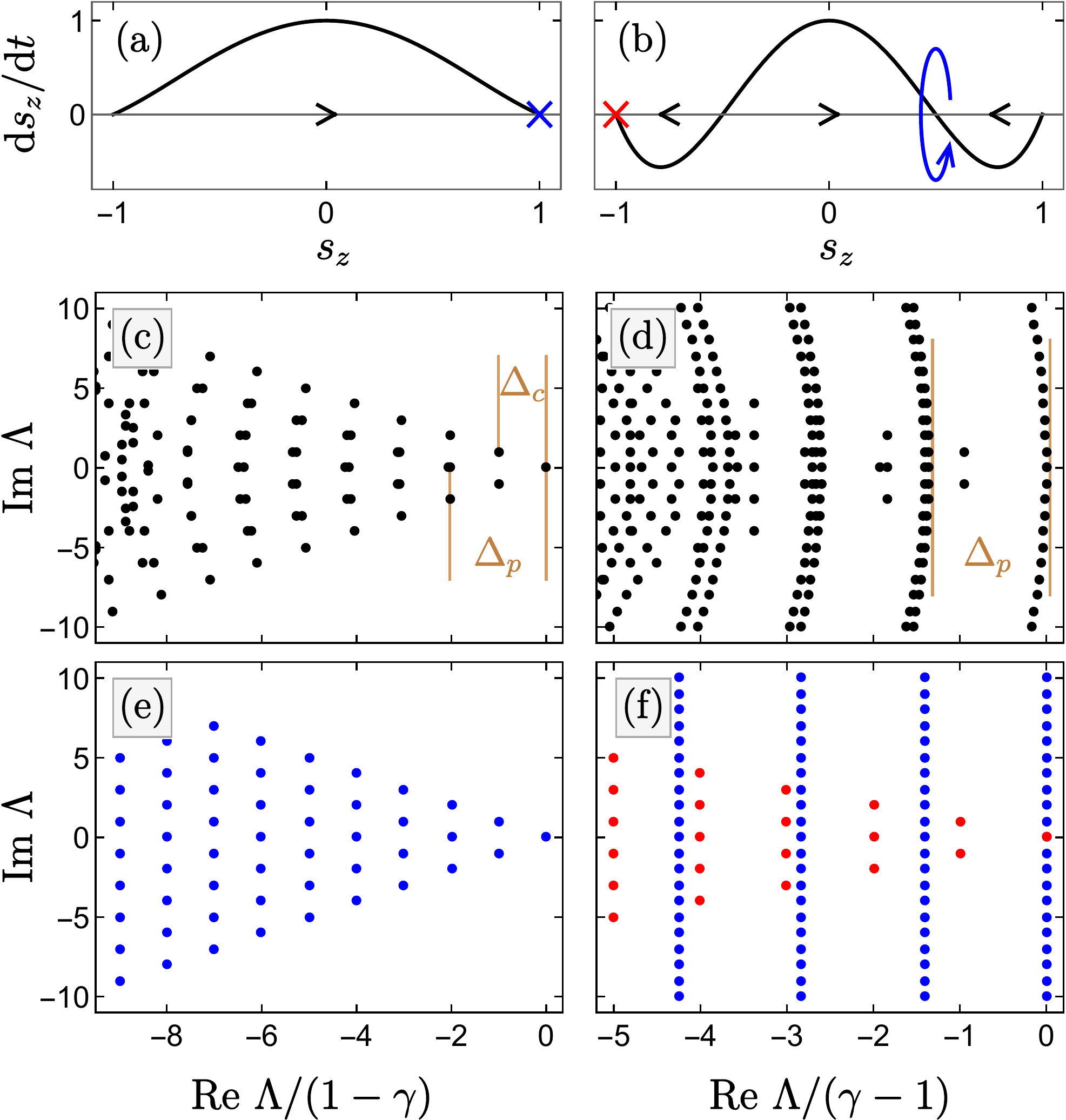}
    \centering
    \caption{\label{fig:spectra_fp_lc}(a),(b) Classical flow  and (c)-(f) quantum spectra for the fixed point (left panels: $\gamma=0.5<1$) and limit cycle (right panels: $\gamma=2>1$) regimes of the spin model.  (a),(b) Crosses are fixed points and curved arrow denotes limit cycle.  (c),(d) Liouvillian spectra for $S=300$.  (e),(f) $S\to\infty$ spectrum from the semiclassical Fokker-Planck equation. In both regimes the gap $\Delta_p$ remains nonzero for $S\to\infty$, signifying the classical approach rate, whereas the gap $\Delta_c$ (decoherence rate) is nonzero for the fixed point  but vanishes for the limit cycle. The blue and red dots in (f) correspond to the limit cycle and the coexisting fixed point at $s_z = -1$, respectively.
    }
\end{figure}

On the other side of the bifurcation, classical trajectories approach a stable fixed point. The quantum signature is a wedge-shaped array of Liouvillian eigenvalues [Figs.\ \ref{fig:spectra_fp_lc}(c) and  \ref{fig:spectra_fp_lc}(e)] that follow from the classical Jacobian.

\emph{Spin model.}|We analyze a spin subject to a Zeeman field $\hat{H}=-\hat{S}_z$ and two competing quantum jump operators, $\hat{L}_1 =  \hat{S}_+ / \sqrt{S}$ and $\hat{L}_2 = \sqrt{\gamma/S^3} \hat{S}_-\hat{S}_z$.  The former relaxes the spin toward the ground state at the north pole.  The latter excites the spin away from it with a rate dependent on $S_z$; this nonlinearity is parametrized by the rate $\gamma$. The normalizations of $\hat{L}_j$ with factors of the spin length $S$ guarantee  consistency in the classical limit $S \to \infty$. In this limit, the Lindblad time evolution of the spin operators reduces to the following equations of motion on the unit (Bloch) sphere \cite{suppl}:
\begin{equation}
    \frac{{\rm d} s_z}{{\rm d} t} = 
    \big(1 - \gamma s_z^2 \big) \big( 1 - s_z^2 \big), \quad
    \frac{{\rm d} \phi}{{\rm d} t} = -1 \;,
    \label{eq:SmSz_classical}
\end{equation}
where $s_z := \langle\hat{S}_z\rangle/S = \cos\theta$ and $\phi :=\arctan \big[\langle\hat{S}_y\rangle / \langle\hat{S}_x\rangle \big]$  can be regarded as classical variables. For $\gamma \leq 1$ all trajectories flow to a unique stable fixed point $s_z = 1$, whereas for $\gamma > 1$ a stable limit cycle at $s_z = 1/\sqrt{\gamma}$ coexists with a stable fixed point at $s_z = -1$, as shown in Figs.\ \ref{fig:spectra_fp_lc}(a) and \ref{fig:spectra_fp_lc}(b). At long times deviation from the final state decays exponentially for any $\gamma \neq 1$. The separation point $\gamma = 1$ features a Hopf bifurcation, at which the decay is algebraic, $s_z \approx 1 - 1/(4t)$.

We now analyze the spectrum and dynamics of the quantum model, explaining how the classical dynamics is recovered. We will later argue that the major findings are generic (model independent). In addition to numerical solutions of the Liouvillian at finite $S$,  we obtain complementary insights using a semiclassical limit of the Lindblad master equation \cite{strunz1998classical, dubois2021semi}.  The resulting Fokker-Planck equation for the phase-space distribution yields the exact spectrum of $\mathcal{L}$ in the $S\to\infty$ limit. For finite $S$ it describes a wave packet that drifts along a classical trajectory and diffuses under quantum fluctuations.

\begin{figure}
    \includegraphics[width=1\columnwidth]{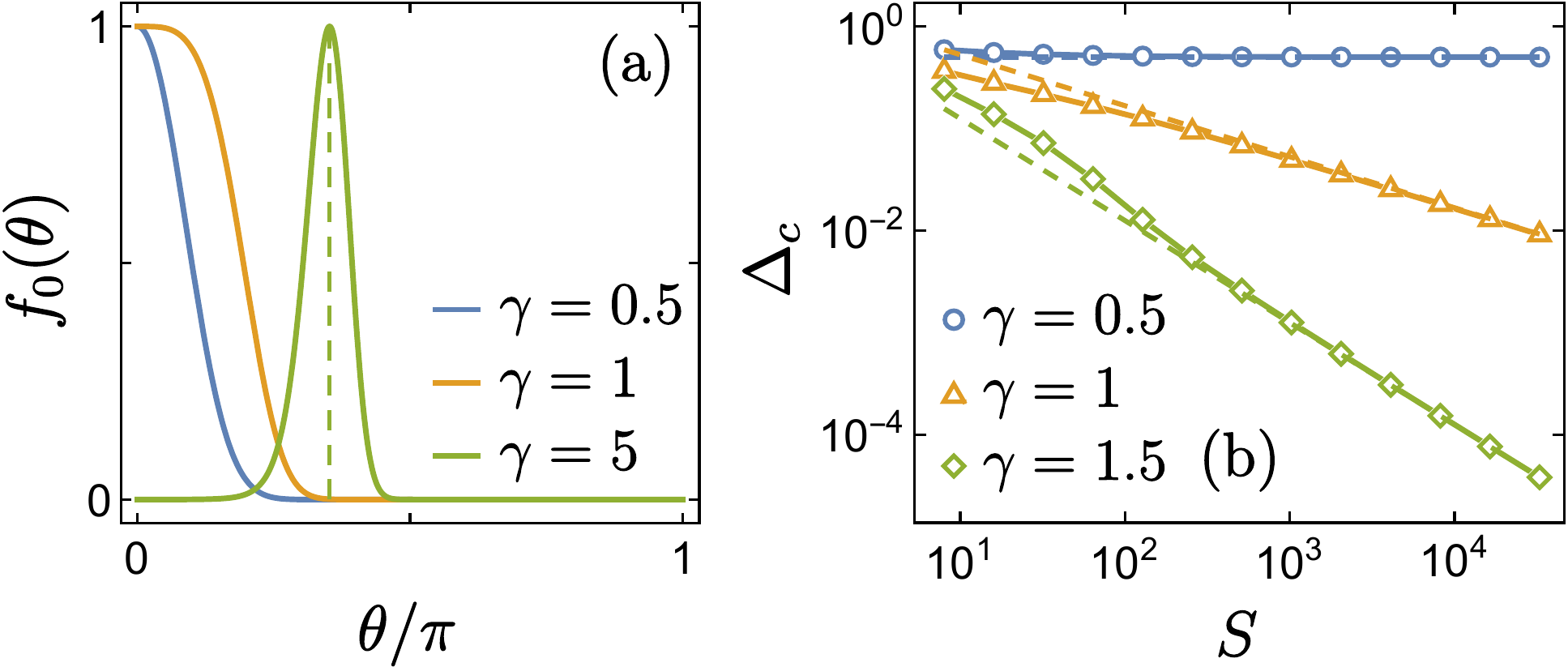}
    \centering
    \caption{(a) Steady-state quasiprobability distribution from the Fokker-Planck equation with $S=20$. It is uniform along $\phi$ and peaked at $\theta=0$ for $\gamma \leq 1$ and at the limit cycle (dashed line) for $\gamma > 1$. Their widths scale as $\Delta\theta \sim \smash{1/\sqrt{S|1-\gamma|}}$ for $\gamma \neq 1$ and $\Delta\theta \sim \smash{S^{-1/4}}$ for $\gamma=1$. (b) The decoherence rate $\Delta_c$ scales as $\smash{S^0}$ for $\gamma<1$ (fixed point), $\smash{S^{-1/2}}$ for $\gamma=1$ (bifurcation point) and $\smash{S^{-1}}$ for $\gamma>1$ (limit cycle). 
    }
    \label{fig:decoherence_rate}
\end{figure}

\emph{Fixed-point regime.}|For $\gamma<1$  the stable fixed point corresponds to the quantum steady state with eigenvalue $\Lambda = 0$ [Figs.~\ref{fig:spectra_fp_lc}(c) and \ref{fig:spectra_fp_lc}(e)].  The steady state for finite $S$ has a distribution of width $\Delta \theta \sim{S}^{-1/2}$ centered at the classical fixed point [Fig.~\ref{fig:decoherence_rate}(a)]. The rest of the spectrum is separated by a minimum real gap of $\Delta_c$ that approaches the classical decay rate $1-\gamma$ as $S \to\ \infty$.

In fact, in the classical limit the low-lying (small-$|\Re \Lambda|$) spectrum, which governs the late-time dynamics, is fully determined by the classical attractor. As shown in Fig.~\ref{fig:spectra_fp_lc}(e), the eigenvalues have the form $\sum_{j = 1,2} n_j\lambda_j$, where $n_j$ are non-negative integers and $\lambda_j$ are eigenvalues of the Jacobian matrix describing the linearized classical dynamics about the north pole [here $\lambda_j = -(1-\lambda) \pm {\rm i}$]. As we discuss later, this structure arises whenever the classical phase space has a stable fixed point. The corresponding eigenstates describe wave packets at different distances from the fixed point and are given by Laguerre polynomials for the spin model \cite{suppl}. Here the wedge-shaped spectrum remains prominent for finite $S$ [Fig.~\ref{fig:spectra_fp_lc}(c)].

\emph{Emergence of limit cycle.}|For $\gamma>1$ the classical limit features an infinite number of equally spaced imaginary eigenvalues.  
The $\Lambda = 0$ state is spread uniformly along the limit cycle with a width $\Delta \theta \sim{S}^{-1/2}$ [Fig.~\ref{fig:decoherence_rate}(a)].  
The other eigenstates of this branch are also clustered along the limit cycle, but with an additional phase winding $e^{{\rm i}l\phi}$. (The index $l$ is the imaginary part of the eigenvalue and is a quantum number for this model \cite{suppl}.)

For any finite $S$ this branch is parabolic,  $\Lambda_l \approx {\rm i}l-(l^2/S)(\gamma^2 - 4\gamma + 5)/(2\gamma-2)$ \cite{suppl},  so there is only a single state with a vanishing decay rate [Fig.~\ref{fig:spectra_fp_lc}(d)].  The curvature can be characterized by the real part $\Delta_c$ of the $l=1$ member of the branch, which collapses as ${S}^{-1}$ [Fig.~\ref{fig:decoherence_rate}(b)].  We argue later that the parabolic shape is responsible for diffusive broadening along the limit cycle as $\sqrt{t/S}$, which is suppressed for $S \to \infty$.

Besides the main branch, there is a series of parallel branches with decay rates $\approx n\Delta_p$ that are also distorted parabolically for finite $S$ [Figs.~\ref{fig:spectra_fp_lc}(d) and \ref{fig:spectra_fp_lc}(f)]. These eigenfunctions (with Hermite polynomial forms \cite{suppl}) support wave packets at increasing distances from the limit cycle, and thus describe approach to the limit cycle at a rate $\Delta_p$, which has the classical value $2(\gamma-1)/\sqrt{\gamma}$ for $S \to \infty$ [Fig.~\ref{fig:hopf_relrate}(c)].

\begin{figure}[tbp]
    \centering
    \includegraphics[width=1\columnwidth]{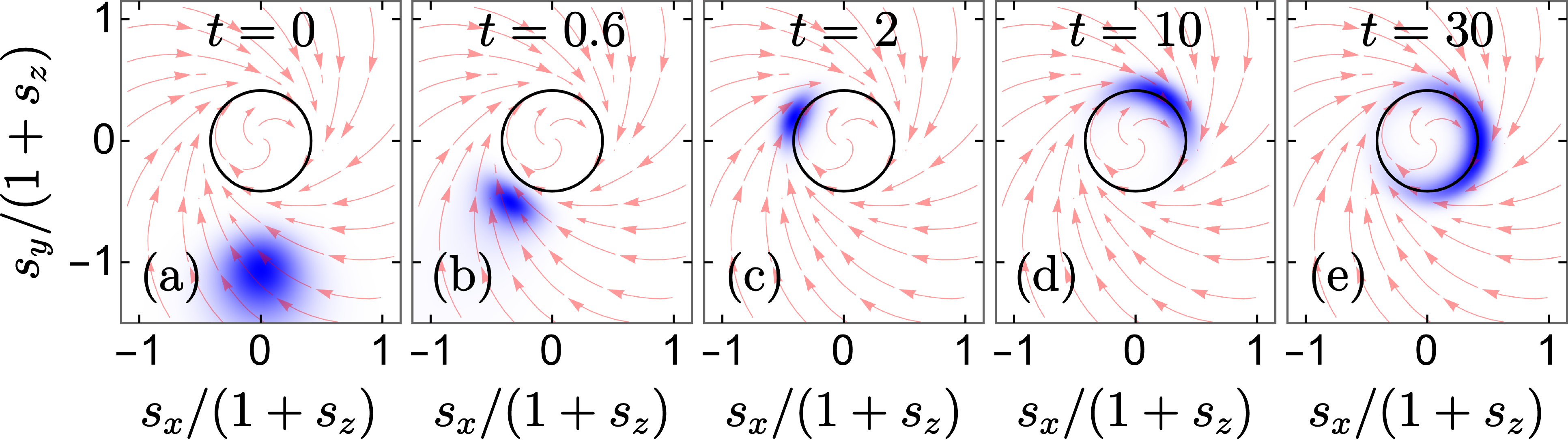}
    \caption{\label{fig:wavepacketdynamics_limitcycle}(a)-(e) Dynamics of quasiprobability distribution in phase space from the Fokker-Planck equation with $S=50$.  The distribution is initially localized at the equator.   Besides approaching the limit cycle as in the classical case, the quantum dynamics shows spreading along the limit cycle (dephasing).
    }
\end{figure}

\begin{figure}[b]
    \includegraphics[width=1\columnwidth]{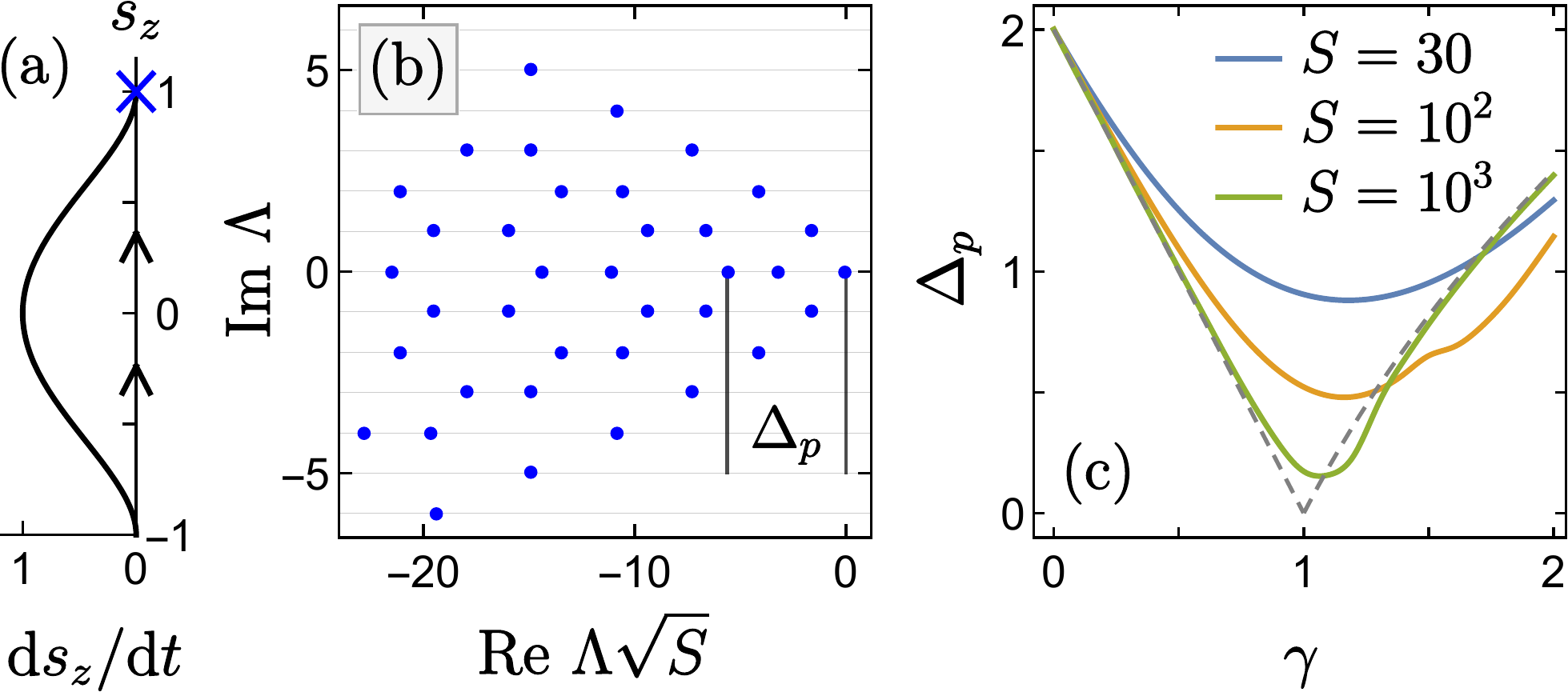}
    \centering
    \caption{(a) Classical flow and (b) the Lindblad spectrum for $S=300$ at the Hopf bifurcation point, $\gamma=1$.  (c) Scaling of the gap $\Delta_p$, defined in Figs.\ \ref{fig:spectra_fp_lc}(c) and \ref{fig:spectra_fp_lc}(d), showing ``critical slowing down'' at $\gamma=1$.
    }
    \label{fig:hopf_relrate}
\end{figure}

Figure \ref{fig:wavepacketdynamics_limitcycle} shows the quantum dynamics for $S=50$ of an initially Gaussian quasiprobability distribution.  The center of mass closely follows the classical trajectories as the wave packet approaches and orbits the limit cycle and then slowly diffuses around it  \cite{Savage_PRA1988_oscillations}, losing phase coherence in the steady state due to the finite value of $S$.

\emph{Fixed point at south pole.}|The classical dynamics \eqref{eq:SmSz_classical} for $\gamma>1$ has a fixed point at $s_z=-1$ (south pole of the Bloch sphere), in addition to the limit cycle.  In the Lindblad spectrum, we see this as an eigenvalue approaching $0$ exponentially with $S$ \cite{suppl}, such that there are two degenerate steady states for $S\to\infty$. [In Fig.~\ref{fig:spectra_fp_lc}(d) this state is too close to distinguish for $S=300$ from the true steady state.]  Furthermore, there is a whole set of eigenvalues governed by this fixed point which follow the $n_1\lambda_1+n_2\lambda_2$ pattern characteristic of fixed points. 

\emph{Bifurcation point: Emergence of algebraic decay.}|In Fig.\ \ref{fig:hopf_relrate}(b) we show the low-lying spectrum at the Hopf bifurcation point $\gamma=1$.  These eigenvalues all collapse onto $\Lambda = {\rm i} l$ with decay rates $\sim S^{-1/2}$.  Infinitely many eigenvalues become degenerate for $S\to\infty$ at every $l$, in contrast to the limit-cycle case where a single branch reaches the imaginary axis.  Thus, for the same harmonic we get a superposition of an infinite number of eigenstates with different radial structures.  It is this combination that allows an algebraic decay to the classical attractor. We explain this later in more detail for a generic (model-independent) setting. The width of the eigenstates scales as $S^{-1/4}$, as opposed to $S^{-1/2}$ for $\gamma \neq 1$ [Fig.~\ref{fig:decoherence_rate}(a)].

\begin{figure}
    \includegraphics[width=1\columnwidth]{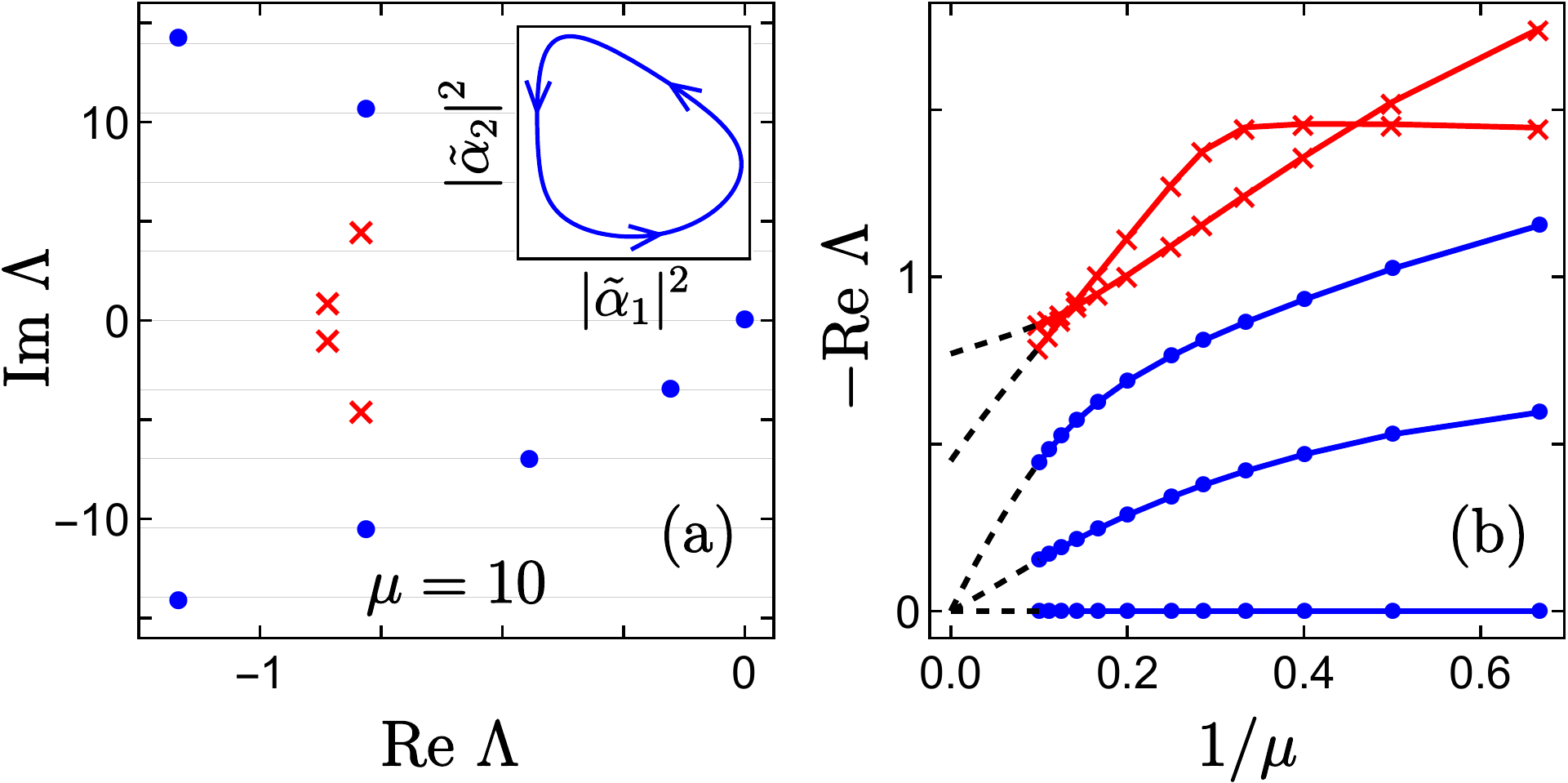}
    \centering
    \caption{Bose-Hubbard dimer in the limit-cycle regime. (a) Low-lying part of spectrum. Inset: the limit cycle projected onto a plane in the four-dimensional phase space, where $|\Tilde{\alpha}_{1,2}|^2$ are rescaled boson numbers in the two sites \cite{suppl}. (b) In the classical limit $\mu\to\infty$ one branch (blue dots) has a vanishing decay rate, corresponding to the limit cycle, while all other eigenvalues  (red crosses) stay in the left half plane. 
    }
    \label{fig:dimer_lc}
\end{figure}

\emph{Dissipative Bose-Hubbard system.}|To verify the generality of our results, we have examined systems other than the spin model. In Fig.~\ref{fig:dimer_lc} we show spectra and scalings for a dissipative Bose-Hubbard dimer. The model is  that introduced in \cite{giraldo2020driven, giraldo2021chaotic, giraldo2022semiclassical}; parameters are documented in \cite{suppl}. The classical limit has a four-dimensional phase space, and there is no quantum number such as $l$ for the spin model; hence, this system is qualitatively very different. The model has a parameter $\mu$ that controls the approach to the classical limit by tuning the average number of bosons. For a choice of parameters leading to a limit cycle in the classical limit, we observe the same signature of a branch collapsing onto the imaginary axis [Fig.\ \ref{fig:dimer_lc}(b)]. The branch follows a parabolic shape more closely as $\mu$ increases.

\emph{Generality.}|We have used specific models to demonstrate spectral signatures for three types of emergent classical behaviors (fixed points, limit cycles, and Hopf bifurcations). We now provide arguments and proofs that these signatures are generic and not model dependent.

In Figs.~\ref{fig:spectra_fp_lc}(e) and \ref{fig:spectra_fp_lc}(f) eigenvalues corresponding to classical fixed points are given by $\Lambda = n_1 \lambda_1 + n_2 \lambda_2$, where $n_1,n_2 = 0,1,2,\dots$ and $\lambda_{1,2}$ are eigenvalues of the linearized classical dynamics around the fixed point. This structure originates from the fact that at long times the dynamics occurs close to the fixed point, so the slow-decaying quantum eigenstates are peaked there, becoming infinitely localized in the classical limit [as in Fig.~\ref{fig:decoherence_rate}(a)]. Under such general conditions one can expand the quantum Fokker-Planck equation about the fixed point and keep the lowest-order terms, which gives a linear drift and a constant diffusion \cite{suppl}. This system can be solved using a ladder-operator construction \cite{leen2016eigenfunctions}, yielding the eigenvalues $\Lambda = \sum_j n_j \lambda_j$, where $n_j = 0,1,2,\dots$ and $\lambda_j$ are eigenvalues of the classical Jacobian. Note that the same spectral form arises for quadratic Lindbladians \cite{prosen2010quantization}.

The classical limit cycle is signaled by a branch that is purely imaginary and equally spaced, where each eigenstate is localized in the $r$ direction and has a different $\phi$ harmonic, $e^{{\rm i}l\phi}$.  This spectral structure is \emph{necessary} to give rise to classical limit cycles.  The azimuthal part of a quantum wave packet can be expanded as $g(\phi,t=0)=\sum_l c_l e^{-{\rm i}l\phi}$, with $l\in\mathbb{Z}$.  For cycling dynamics,  $g(\phi,t) = g(\phi-\omega{t},0)=  \sum_l c_l e^{-{\rm i}l\phi}e^{{\rm i}l\omega{t}}$.  Since a state with eigenvalue $\Lambda$ contributes a factor $e^{\Lambda t}$ to the dynamics, this implies that the eigenvalues are $\Lambda_l={\rm i}l\omega$ and the eigenfunctions have the angular form $\sim{e}^{-{\rm i}l\phi}$. To reproduce a sharp point on a classical trajectory, all Fourier modes (each $l\in\mathbb{Z}$) must be present.

Close to the classical limit the branch is parabolic.  This leads to $g(\phi,t)= \sum_l c_l e^{-{\rm i}l(\phi-\omega{t})}e^{-l^2t/\tau} \sim e^{\tau(\phi-\omega{t})^2/4t}$, where $\tau \to \infty$ in the classical limit. Hence, the wave packet broadens as $\Delta\phi\sim\sqrt{t/\tau}$.  Such diffusive spreading, expected from the diffusion term in the Fokker-Planck equation, is thus coupled to the parabolic distortion of the branch, which should be generic.

Close to the limit cycle a classical trajectory has a vanishing radial speed, $\dot{r}\to0$, but a nonzero angular speed, $r\dot{\phi}\to r_{\text{lc}}\dot{\phi}$. This decoupling of timescales produces many quasistationary (slow-decaying) orbits at small distances from the limit cycle, which show up in the spectrum as additional branches. Note that these branches are absent for a fixed point where the radial and angular speeds decay proportionally.

Finally, we reported above that the classical bifurcation point involves many quantum eigenvalues corresponding to the same harmonic, and hence different $r$ structures, collapsing onto the imaginary axis. This enables the characteristic algebraic decay, as 
$\xpct{r} = \sum_n c_n \xpct{r}_n e^{-\xi_n t/\sqrt{S}}$ is now obtained as an infinite sum over vanishing decay rates ($n=0,1,2,\ldots$). Here, $c_n$ define the initial state in terms of eigenstate overlaps, $\xpct{r}_n$ are the eigenstate expectation values of the radial coordinate, and $-\xi_n/\sqrt{S}$ are the eigenvalues. (The values of $c_n$ and $\xpct{r}_n$ depend on the normalization of the left and right eigenstates, but their product is uniquely defined.) In general, power-law scalings of these quantities with $n$ can give rise to algebraic decay; the specific scalings are likely system dependent.  In the spin model, $\xi_n\sim n^{3/2}$, and for a localized initial state, this leads to $\xpct{r}\sim t^{-1/2}$, as detailed in the Supplemental Material \cite{suppl}.  

For a Hopf bifurcation, generically $\dot{r}/(r\dot{\phi})\sim r$ vanishes close to the classical attractor.  Thus, we again have a dynamical decoupling of radial and angular motion. The quasistationary orbits at small $r$ have the same angular frequency, which implies that the imaginary parts of the low-lying eigenvalues are equally spaced.

A generic system might be expected to have several fixed points and/or limit cycles.  Our observation for the spin system at $\gamma>1$ [Fig.~\ref{fig:spectra_fp_lc}(f)] suggests that each such feature governs a collection of eigenvalues in the corresponding quantum Liouvillian spectrum.

\emph{Context and discussion.}|We have initiated the study of the spectral origin of nonlinear dynamical phenomena in the classical limit of quantum dissipative physics.  Such classical limits are useful for understanding physical phenomena in different setups \cite{Savage_PRA1988_oscillations, strunz1998classical, ArmenMabuchi_PRA2006_CavityQED_bifurcations,  Bhaseen_Keeling_PRA2012_Dicke, LeeSadeghpour_PRL2013_quantumVanDerPolSynchronization, Marquardt_Clerk_PRX2014, Lee_Chan_Wang_PRE2014_synchronization, NunnenkampBruder_AnnPhys2015_quantumVanDerPolSynchronization, Bruder_PRL2018_synchronization_1spin, Poletti_PeriodDoubling_2018, Ferreira_Ribeiro_PRB2019_dissipativeLMG, Dutta_Cooper_PRL2019, Mok_Heimonen_PRR2020_synchronization, FernengelDrossel_JPA2020_BifurcationsChaos, Huybrechts_Minganti_Nori_Shammah_PRB2020, Arosh_Cross_Lifshitz_PRR2021_limitcycles_Rayleigh_VanDerPol, Thomas_Sentilvelan_PRA2022_synchr, stitely2020nonlinear, Seibold_Savona_PRA2020_timecrystal, dubois2021semi, Kato_Nakao_SciRep2022_TuringInstability,  LiFazioChesi_NJP2022_semiclassicalDicke, Bakker_Gritsev_Krimer_PRL2022, Zhang_etal_PRR2023_quantumsynchr,cosme2022observation, LiChesi_PRA2024_DickeChaos}. We have elucidated the prototypical cases of fixed points, limit cycles, and critical slowing down. 
While such phenomena have been predicted in quantum systems using semiclassical equations of motion, this Letter provides a foundation for how they arise from the full quantum generator.

Algebraic decay has previously been seen for the thermodynamic limit \cite{Tomadin_Diehl_Zoller_PRA2011, Cai_Barthel_PRL2013_algebraic, Znidaric_PRE2015_relaxationtimes, Medvedyeva_Kehrein_PRB2014, Vincenzi_Minganti_Orso_Ciuti_PRA2018_CriticalSlowingDown_BHlattices, Minganti_Nori_PRR2021_SpectralCollapse_ScullyLamb}.  Here, we illustrate the general mechanism of algebraic decay in the classical limit, as also observed in Ref.\ \cite{Poletti_Bernier_Georges_Kollath_PRL2012}.  
Our findings regarding the spectral signatures of a limit cycle should also apply to other open quantum systems that support limit cycles in the classical limit, e.g., a quantum Van der Pol oscillator \cite{Dutta_Cooper_PRL2019, Mok_Heimonen_PRR2020_synchronization, Arosh_Cross_Lifshitz_PRR2021_limitcycles_Rayleigh_VanDerPol, Thomas_Sentilvelan_PRA2022_synchr, Cabot_Giorgi_Zambrini_2023nonequilibrium}, an open Dicke model \cite{Bhaseen_Keeling_PRA2012_Dicke, stitely2020nonlinear, LiFazioChesi_NJP2022_semiclassicalDicke, Nie_Zheng_PRA2023_Dicke}, and others \cite{Chan_Lee_Gopalakrishnan_PRA2015, Piazza_Ritsch_PRL2015, Cosme_2024_limitcycle}. We have shown how the apporach of an infinite number of eigenvalues to the imaginary axis provides mechanisms for both limit-cycle dynamics and algebraic decay in the classical limit. Liouvillian eigenvalues located on or approaching the imaginary axis have also been studied from other perspectives \cite{Baumgartner_Narnhofer_JPA2008,  Albert_Jiang_PRA2014_Lindbladsymmetries, AlbertEtAl_PRX2016_geometryresponse, KeelingFazio_PRL2018_BoundaryTimeCrystals, BucaTindallJaksch_NatComm2019, BucaJaksch_PRL2019, Seibold_Savona_PRA2020_timecrystal, BookerBucaJaksch_NJP2020, Prazeres_Iemeni_PRB2021_BoundaryTimeCrystals, Minganti_Nori_PRR2021_SpectralCollapse_ScullyLamb, BucaBookerJaksch_SciPost2022_synchronization,  Alaeian_Buca_CommPhys2022, Seibold_Savona_PRA2022_dissipativeKerrSolitons, Dubois_Saalman_Rost_PRR2023, Souza_dosPazeres_Iemini_PRL2023_sufficient, Tindall_Jaksch_Munoz_SciPost2023, LiLiJin_PRA2023_synchron_DecFreeSubspace, LiLiJin_PRA2023_QuantumNonstationary, Nakanishi_Sasamoto_PRA2023, Krishna_Vinjanampathy_PRL2023, Iemini_Chang_Marino_PRA2024}.

Our results open up several research directions and questions: (1)  It remains to be explained how other classical nonlinear phenomena, particularly period doubling to chaos, emerges from Liouvillian spectra \cite{LiChesi_PRA2024_DickeChaos}.  (2) How do the spectra of Liouvillian maps (confined to the unit circle rather than the negative half plane) lead to discrete-time nonlinear phenomena \cite{Poletti_PeriodDoubling_2018}?  (3) Can one connect  classical nonlinear phenomena to statistical aspects of the  Liouvillian spectra, analogous to the 
Bohigas-Giannoni-Schmit conjecture \cite{Bohigas_PRL1984} for the Hamiltonian case?

\emph{Acknowledgments.}|We thank Nigel Cooper, Jonathan Dubois, and Felix Fritzsch for useful discussions. 
This work was supported in part by the Deutsche Forschungsgemeinschaft through SFB 1143 (project ID 247310070). 

\begingroup
\renewcommand{\addcontentsline}[3]{}
\renewcommand{\section}[2]{}

\endgroup

\onecolumngrid
\clearpage
\renewcommand{\baselinestretch}{1.3}\normalsize
\begin{center}
\textbf{\large Supplemental Material for\\
	    \emph{Quantum Origin of Limit Cycles, Fixed Points, and Critical Slowing Down}}\\
\vspace{0.5cm}
Shovan Dutta, Shu Zhang, and Masudul Haque
\end{center}
\vspace{1cm}



    \setcounter{table}{0}
    \renewcommand{\thetable}{S\arabic{table}}%
    \setcounter{figure}{0}
    \renewcommand{\thefigure}{S\arabic{figure}}%
    \renewcommand{\theHfigure}{S\arabic{figure}}
    \setcounter{secnumdepth}{3}
    \setcounter{section}{0}
    \renewcommand{\thesection}{S\arabic{section}}%
    \setcounter{subsection}{0}
    \renewcommand{\thesubsection}{S\arabic{section}.\arabic{subsection}}
    \setcounter{equation}{0}
    \renewcommand{\theequation}{S\arabic{equation}}%
    \setcounter{page}{1}
    \renewcommand{\thepage}{SM-\arabic{page}}


\makeatletter
\renewcommand{\baselinestretch}{1}\normalsize
\tableofcontents

\section{Overview}

\begin{itemize}[leftmargin=*]

\item We briefly review the Lindblad (GKSL) formalism and the Fokker-Planck formalism that emerges in the classical limit.  (Sec.~\ref{suppsec:GKSL_FokkerPlanck}.)

\item We describe the classical dynamics of the spin model used in the main text, providing some further details and some pictorial representations, in   Sec.~\ref{suppsec:spinmodel_classical}.

\item We explain a `diagonal' structure in the spin model: each sub- and super-diagonal evolves independently under Lindblad dynamics.   (Sec.~\ref{suppsec:spin_decoupling}.)
 This structure results in the quantum number $l$ introduced and used in the main text.  This quantum number labels the imaginary parts of the eigenvalues.  

\item We outline how the spectrum is obtained using the semiclassical Fokker-Planck formalism. (Sec.~\ref{suppsec:spinmodel_semiclassical_spectrum}.)

\item We show the animations of the Fokker-Planck dynamics of the spin model on the Bloch sphere. (Sec.~\ref{suppsec:spinmodel_animations}.)

\item In the main text (in addition to the primary results on Hopf bifurcations and limit cycles) we also have presented a result on the quantum Liouvillian spectrum corresponding to \emph{fixed points}:  a universal wedge-shaped arrangement of eigenvalues.  This is derived and discussed in Sec.~\ref{suppsec:fixedpoint_wedge}.

\item The spectral signature of limit cycles (parabolic branches which become fully vertical in the classical limit) can be derived using non-Hermitian perturbation theory.  Sec.~\ref{suppsec:limtcyclebranch_perturbthy} provides details for the spin model.  

\item One of our main results is the mechanism by which algebraic relaxation emerges at a critical (Hopf bifurcation) point.  In the main text we have explained the mechanism qualitatively and generally.  In Sec.~\ref{suppsec:algebraic_decay} we describe the specifics of this mechanism for the particular case of the nonlinear spin model.  

\item In Sec.~\ref{suppsec:BHdimer}, we provide some details of our work with the driven-dissipative Bose-Hubbard dynamics.  As this is a very different system compared to the nonlinear spin model, it serves to highlight the generality of our results.  
    
\end{itemize}

\section{Lindblad (GKSL) description and Fokker-Planck formalism \label{suppsec:GKSL_FokkerPlanck}}

In the widely used Gorini–Kossakowski–Sudarshan–Lindblad (GKSL) formalism \cite{supp_GKSL_1976, supp_Lindblad1976}, the time evolution of the density matrix is given by 
\begin{equation}
    \frac{{\rm d}}{{\rm d}t} \rho = -{\rm i} [\hat{H}, \rho] + \frac{1}{2} \sum_k \left( [\hat{L}_k \rho, \hat{L}_k^\dagger]
    + [\hat{L}_k , \rho \hat{L}_k^\dagger]\right) ,
    \label{eq:lindblad}
\end{equation}
where the unitary dynamics of the system is described by the Hamiltonian $\hat{H}$ and the dissipative dynamics is described by a set of jump operators $\{\hat{L}_k\}$.  

In the semiclassical limit [up to $O(\hbar^2)$], this master equation reduces to a Fokker-Planck equation for a quasiprobability distribution $\rho(\mathbf{z},t)$ in phase space \cite{supp_strunz1998classical, supp_dubois2021semi}, namely
\begin{equation}
    \frac{\partial \rho}{\partial t} 
    + \sum_i \frac{\partial}{\partial z_i} [u_i(\mathbf{z}) \rho] - \frac{1}{2} \sum_{i,j} \frac{\partial^2}{\partial z_i \partial z_j} [D_{i,j}(\mathbf{z}) \rho] 
    = 0 \;,
    \label{eq:fokkerplanck}
\end{equation}
where $z_i$ are the phase-space coordinates, e.g., $(x,p)$ for an oscillator or $(S_x,S_y,S_z)$ for a spin. The objects
\begin{equation}
    u_i =\; \{z_i, H\} + \sum_k \text{Im } L_k \{z_i, L_k^*\} 
    + \frac{1}{2} \sum_k \text{Re } \{\{z_i, L_k^*\}, L_k\} 
    \label{eq:driftvec}
\end{equation}    
and
\begin{equation}
    D_{i,j} =\; \sum_k \text{Re } \{z_i, L_k\} \{z_j, L_k^*\} 
    \label{eq:diffusionmat}    
\end{equation}
are the drift vector and the diffusion matrix, respectively.
Here, $H$ and $L_k$ are functions of $\mathbf{z}$ obtained as the classical limit of the Hamiltonian operator and the Lindblad jump operators, respectively.  Also,  $\{F,G\}$ denotes the Poisson bracket of $F$ and $G$, which result from the commutators. For canonical coordinates $\mathbf{z} = (\mathbf{x}, \mathbf{p})$,
\begin{equation}
    \{F,G\} =  \frac{\partial F}{\partial \mathbf{x}} \frac{\partial G}{\partial \mathbf{p}} - \frac{\partial F}{\partial \mathbf{p}} \frac{\partial G}{\partial \mathbf{x}},
\end{equation} 
whereas for a spin $\mathbf{z} = (S_x, S_y, S_z)$,
\begin{equation}
    \{F,G\} = \mathbf{z} \cdot \left( \frac{\partial F}{\partial \mathbf{z}} \times \frac{\partial G}{\partial \mathbf{z}} \right) .
\end{equation} 
The expectation of an observable $F(\mathbf{z})$ is given by $\langle F \rangle = \int d\mathbf{z} F(\mathbf{z}) \rho(\mathbf{z},t)$. From Eq.~\eqref{eq:fokkerplanck} one finds
\begin{equation}
    \frac{{\rm d}}{{\rm d}t} \langle F \rangle =
    \left\langle 
    \sum_{i} \frac{\partial F}{\partial z_i} u_i (\mathbf{z},t) 
    + \frac{1}{2} \sum_{i,j} \frac{\partial^2 F}{\partial z_i \partial z_j} D_{i,j} \!
    \right\rangle \;.
\end{equation}
In particular, the average coordinates are governed by the drift, ${\rm d}\langle\mathbf{z}\rangle/{\rm d}t = \langle u_{\mathbf{z}} \rangle$, which follows a classical trajectory for a localized wavepacket.

\section{Spin model: Classical behavior \label{suppsec:spinmodel_classical}}

For a quantum system described by a Hamiltonian $H$ and Lindblad operators $\{ \hat{L}_k \}$ the Heisenberg equation of motion for an operator $\hat{X}$ can be found from Eq.~\eqref{eq:lindblad}, which yields
\begin{equation}
    \frac{{\rm d} \hat{X}}{{\rm d} t} = 
    {\rm i} [\hat{H}, \hat{X}] + \frac{1}{2} \sum_k \Big( \hat{L}_k^{\dagger} [\hat{X}, \hat{L}_k] +
    [\hat{L}_k^{\dagger}, \hat{X}] \hat{L}_k \Big) \;.
    \label{eq:lindblad_heisenberg_eom}
\end{equation}

For our single-spin system subject to the Hamiltonian $\hat{H}= -\hat{S}_z$ and the jump operators $\hat{L}_1 =  \hat{S}_+ / \sqrt{S} $ and $\hat{L}_2 = \sqrt{\gamma/S^3} \hat{S}_-\hat{S}_z$, 
the spin operators evolve as follows:
\begin{align}
    \frac{{\rm d} \hat{S}_z}{{\rm d} t} & =
    \hat{S}_- \hat{S}_+ / S
    -\left( \gamma/S^3 \right) \hat{S}_+ \hat{S}_- \hat{S}_z^2 , \\
    \frac{{\rm d} \hat{S}_+}{{\rm d} t} & = 
    -{\rm i} \hat{S}_+ - \hat{S}_z \hat{S}_+ 
    + \left( \gamma/S^3 \right) \hat{S}_+ \left(\hat{S}_z^3 + 2\hat{S}_z^2 - \frac{1}{2} \hat{S}_- \hat{S}_+ \right) , \\
    \frac{{\rm d} \hat{S}_-}{{\rm d} t} & =
    {\rm i} \hat{S}_- - \hat{S}_- \hat{S}_z
    + \left( \gamma/S^3 \right)  \left(\hat{S}_z^3 + 2 \hat{S}_z^2 - \frac{1}{2} \hat{S}_- \hat{S}_+ \right) \hat{S}_-  .
\end{align}
With $s_z := S_z/S$ and $s_\pm := S_\pm/S $, these equations give, up to $O(1/S)$,
\begin{align}\label{eq:classical-eom}
    \frac{{\rm d} s_z}{{\rm d} t} & =
    \left( 1 - \gamma s_z^2 \right) \left( 1 - s_z^2 \right) , \\
    \frac{{\rm d} s_\pm}{{\rm d} t} & = 
    \mp{\rm i} s_\pm - \left( 1 - \gamma s_z^2 \right) s_z s_\pm \;.
\end{align}
This is equivalent to taking the classical limit, where all commutators are neglected. In particular, we have used $\hat{S}_+ \hat{S}_- \approx \hat{S}_- \hat{S}_+ \approx S^2 - \hat{S}_z^2$ for $S \rightarrow \infty$. 

\begin{figure}[!h]\label{fig-classical-trajectory}
\centering
\includegraphics[width=0.7\columnwidth]{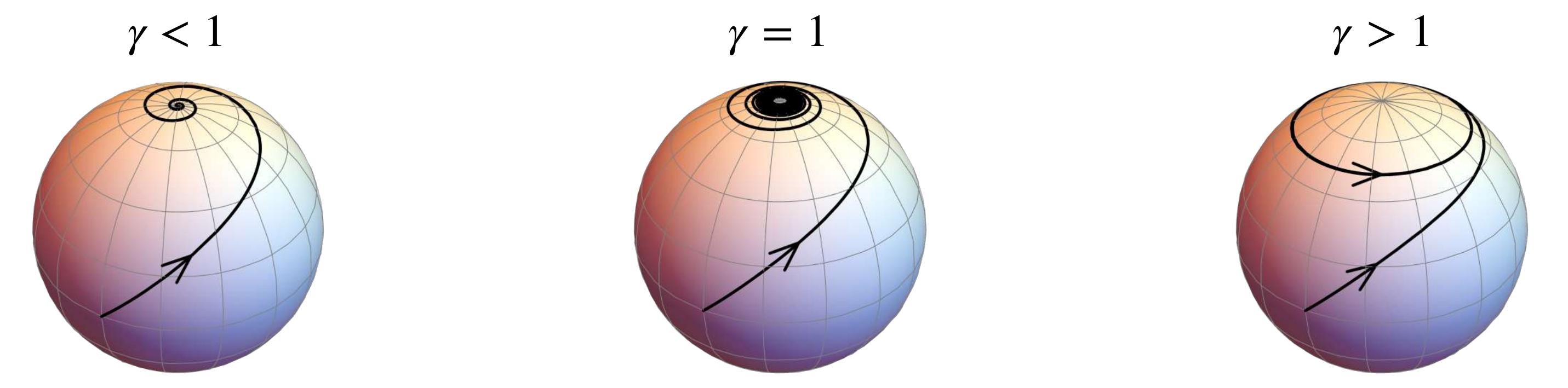}
\caption{\label{fig-classical-trajectory}Classical trajectories exponentially approaching the fixed point for $\gamma<1$, algebraically approaching the north pole at the Hopf bifurcation point  $\gamma=1$, and exponentially approaching the limit cycle for $\gamma>1$. }
\end{figure}

To understand the relaxation to the three classical attractors at $s_z = \pm 1$ and $1/\sqrt{\gamma}$ respectively, we define $\epsilon_{\pm} := 1 \mp s_z$, $\epsilon_{\text{lc}} := 1/\sqrt{\gamma} - s_z$, and obtain
\begin{align}
    \frac{{\rm d}\epsilon_{\pm}}{{\rm d}t} &= 
    \pm 2(\gamma-1) \epsilon_{\pm} \mp (5\gamma-1) \epsilon_{\pm}^2 \pm 4\gamma \epsilon_{\pm}^3 \mp \gamma \epsilon_{\pm}^4 \;,
    \label{eq:SmSz_relaxpm}\\[0.1cm]
    \frac{{\rm d}\epsilon_{\text{lc}}}{{\rm d}t} &= 
    -2 \frac{\gamma-1}{\sqrt{\gamma}} \epsilon_{\text{lc}} 
    + (\gamma-5) \epsilon_{\text{lc}}^2 
    + 4\sqrt{\gamma} \epsilon_{\text{lc}}^3 
    - \gamma \epsilon_{\text{lc}}^4 \;.
    \label{eq:SmSz_relaxlc}
\end{align}
The deviations decay exponentially in time except at the Hopf bifurcation point,  $\gamma=1$, where the linear term on the right-hand side vanishes, leading to a purely algebraic decay given by ${\rm d}\epsilon_+/{\rm d}t \approx -4 \epsilon_+^2$ (see Fig.~\ref{fig-classical-trajectory}).

\section{Spin model: Diagonal decoupling \label{suppsec:spin_decoupling}}

Using the Hamiltonian and jump operators for the spin model in Eq.~\eqref{eq:lindblad} gives the equation of motion for the density-matrix elements $\rho_{m,m'}$,
\begin{align} \label{eq:density-matrix-diagonal}
    \nonumber \frac{{\rm d}}{{\rm d}t} \rho_{m,m'} &= {\rm i}(m-m') \rho_{m,m'} \\
    \nonumber &+ \frac{1}{S} \left[
    \beta_{S,-m} \beta_{S,-m^{\prime}} \; \rho_{m-1,m^{\prime}-1} 
    - \frac{1}{2} \left( \beta_{S,m}^2 + \beta_{S,m^{\prime}}^2 \right) \rho_{m,m^{\prime}}
    \right] \\
    &+ 
    \frac{\gamma}{S^3} \left[
    (m+1)(m^{\prime}+1) \beta_{S,m} \beta_{S,m^{\prime}} \; \rho_{m+1,m^{\prime}+1} 
    - \frac{1}{2} \left( m^2 \beta_{S,-m}^2 + m^{\prime 2} \beta_{S,-m^{\prime}}^2 \right) \rho_{m,m^{\prime}}
    \right] ,
\end{align}
where $\beta_{S,m} := \sqrt{S(S+1)-m(m+1)}$. 
Noticeably, different diagonals of the density matrix decouple, yielding $2S+1$ independent equations for $l = m-m' = 0, 1, \dots 2S$. This is a result of a weak symmetry \cite{supp_Buca_Prosen_NJP2012} corresponding to rotation about the $z$ axis. The imaginary term in Eq.~\eqref{eq:density-matrix-diagonal} can be eliminated by transforming to a rotating frame, $\tilde{\rho}_{m,m'} = \rho_{m,m'} e^{-{\rm i}lt}$. The time evolution of the $2S+1-l$ elements on the $l$-th diagonal of the density matrix consists of a phase modulation at the same angular frequency $l$ and a coupling of the magnitudes among themselves. Their normal modes can be viewed as a series of circular harmonics of the same angular momentum and different radial structures, which will become more evident in the construction below.


\section{Spin model: Spectrum from Fokker-Planck description \label{suppsec:spinmodel_semiclassical_spectrum}}

Using spherical coordinates $\mathbf{z} = S(\sin \theta \cos\phi,\; \sin \theta \sin \phi,\; \cos \theta)$ in Eq.~(\ref{eq:fokkerplanck}), we obtain the Fokker-Planck equation
\begin{align}
    \nonumber
    \frac{\partial \rho}{\partial t} = 
    \frac{\partial \rho}{\partial \phi} 
    &+ \left[ 
    \frac{1}{2S \tan^2\theta} \frac{\partial^2 \rho}{\partial \phi^2} 
    + 2 \cos\theta \rho 
    + \left( \sin\theta + \frac{1}{2S \tan\theta} \right) \frac{\partial \rho}{\partial \theta}
    + \frac{1}{2S} \frac{\partial^2 \rho}{\partial \theta^2} 
    \right] \\[0.1cm]
    &+ \frac{\gamma}{2} \left[ 
    \frac{\cos^2 2\theta}{S \sin^2\theta} \frac{\partial^2 \rho}{\partial \phi^2} 
    - \frac{\sin 4\theta}{\sin\theta} \rho
    - \left( 2 \sin\theta \cos^2 \theta + \frac{1 - 3 \cos 2\theta}{2S \tan\theta} \right) \frac{\partial \rho}{\partial \theta} 
    + \frac{\cos^2 \theta}{S} \frac{\partial^2 \rho}{\partial \theta^2}
    \right] .
    \label{eq:SmSz_fokkerplanck}
\end{align}
This equation has solutions of the form $\rho = f(\theta) e^{ {\rm i} l \phi} e^{\lambda t}$, where $f(\theta)$ satisfies the eigenvalue equation
\begin{equation}
    ( \lambda-  {\rm i} l)  f = g_0(\theta) f + g_1(\theta) \frac{df}{d\theta} + g_2(\theta) \frac{d^2 f}{d\theta^2} \;,
    \label{eq:radial-f}
\end{equation}
with
\begin{align}
    g_0(\theta) &= 2 \cos\theta - \frac{l^2}{2S \tan^2 \theta} - \frac{\gamma}{2} \left( \frac{\sin 4\theta}{\sin\theta} + \frac{l^2}{S} \frac{\cos^2 2\theta}{\sin^2 \theta} \right) \;,
    \label{eq:SmSz_g0}\\[0.1cm]
    g_1(\theta) &= \sin\theta + \frac{1}{2S \tan\theta} - \gamma \left( \sin\theta \cos^2 \theta + \frac{1 - 3 \cos 2\theta}{4S \tan\theta} \right) \;,
    \label{eq:SmSz_g1}\\[0.1cm]
    g_2(\theta) &= \frac{1 + \gamma \cos^2 \theta}{2S} \;.
    \label{eq:SmSz_g2}
\end{align}
For finite $S$ there exists a unique steady state with $\lambda = l=0$,
\begin{equation}
    f_0(\theta) \propto e^{{-2S \left[ \cos\theta - (2/\sqrt{\gamma}) \arctan (\sqrt{\gamma} \cos\theta) \right]}} \;,
    \label{eq:SmSz_steadystate}
\end{equation}
which is peaked at $\theta_c = 0$ for $\gamma \leq 1$, corresponding to the classical fixed point, and at $\theta_c = \arccos(1/\sqrt{\gamma})$ for $\gamma > 1$, corresponding to the classical limit cycle. 
The low-lying spectrum can be found by a $1/S$ expansion of Eq.~(\ref{eq:radial-f}) after a proper rescaling of the coordinates.

\subsection{Fixed point for $\gamma < 1$}

For $\gamma < 1$ the steady state in Eq.~\eqref{eq:SmSz_steadystate} varies as $f_0(\theta) \sim e^{-(\theta/\Delta\theta)^2}$, where $\Delta\theta = S^{-1/2} \sqrt{(1+\gamma)/(1-\gamma)}$. To find the other eigenstates we write $\theta := r \Delta\theta$ in Eq.~\eqref{eq:radial-f} and expand in powers of $1/S$, obtaining
\begin{equation}
    (\lambda - {\rm i}l) f = 
    (1-\gamma) \left[ \frac{1}{2} \frac{{\rm d}^2f}{{\rm d}r^2} + \left(r + \frac{1}{2r} \right) \frac{{\rm d}f}{{\rm d}r} + \left( 2 - \frac{l^2}{2 r^2} \right) f \right] + O(1/S) \;,
    \label{eq:SmSz_dampedeqn}
\end{equation}
which has the solutions $\lambda = {\rm i}l - (1-\gamma) (|l| + 2j) + O(1/S)$ and $f(r) \propto r^{|l|} e^{-r^2} L_j^{|l|}(r^2) + O(1/S)$ for $j = 0, 1, 2, \dots$, where $L_j^{\nu}$ are the generalized Laguerre polynomials.  These eigenvalues form a wedge-shaped pattern, as shown and discussed in the main text, and derived in a more general setting in Sec.~\ref{suppsec:fixedpoint_wedge}.

\subsection{$\gamma > 1$: Limit cycle and additional fixed point}
\label{suppsec:spinmodel_limitcycle_spectrum}

Here, the steady state in Eq.~\eqref{eq:SmSz_steadystate} varies as $f_0(\theta) \sim e^{-x^2}$ where $\theta := \theta_c + x \Delta \theta$ and $\Delta\theta = \gamma^{1/4} / \sqrt{S (\gamma-1)}$. Writing Eq.~\eqref{eq:radial-f} in terms of $x$ we find
\begin{equation}
    (\lambda - {\rm i}l) f = 
    \frac{\gamma-1}{\sqrt{\gamma}} \left[ \frac{{\rm d}^2f}{{\rm d}x^2} + 2 x \frac{{\rm d}f}{{\rm d}x} + 2 f \right] + O(1/\sqrt{S}) \;,
    \label{eq:SmSz_lceqn}
\end{equation}
which gives $\lambda = {\rm i}l -2 [(\gamma-1) / \sqrt{\gamma}] j + O(1/\sqrt{S})$ and $f(x) \propto e^{-x^2} H_j(x) + O(1/\sqrt{S})$ for $j=0,1,2,\dots$, where $H_j$ are the Hermite polynomials. The $j=0$ branch is purely imaginary in the classical limit $\lim_{S\rightarrow \infty} \lambda_{\text{lc}}= {\rm i}l$, forming a Fourier basis for the persistent oscillation around the limit cycle.  Each additional branch ($j=1,2,3,\ldots$) is parallel to the imaginary axis. Classical trajectories approach the limit cycle at a rate determined by the gap to the $j=1$ branch.

In Sec.~\ref{suppsec:limtcyclebranch_perturbthy},  we will extend these results by determining the perturbative corrections to the eigenvalues for finite $S$.  This will yield the parabolic distortion of $O(1/S)$ reported in the main text. (Note that the correction is of second order.)  This shape is responsible for diffusion along the limit cycle.  

For $\gamma > 1$ a second group of eigenstates peak at $\theta'_c = \pi$, which corresponds to the stable fixed point at the south pole in the classical limit. This fixed point gives rise to a low-lying spectrum similar to that of the fixed point at the north pole for $\gamma<1$. Here, $\lambda = {\rm i}l - (\gamma - 1) (|l| + 2j) + O(1/S)$. Due to the presence of two classical steady states|a limit cycle and a fixed point|the eigenstates of the full Fokker-Planck equation [Eq.~\eqref{eq:radial-f}] develop a richer structure.

\begin{figure}[!h]
\centering
\includegraphics[width=0.7\columnwidth]{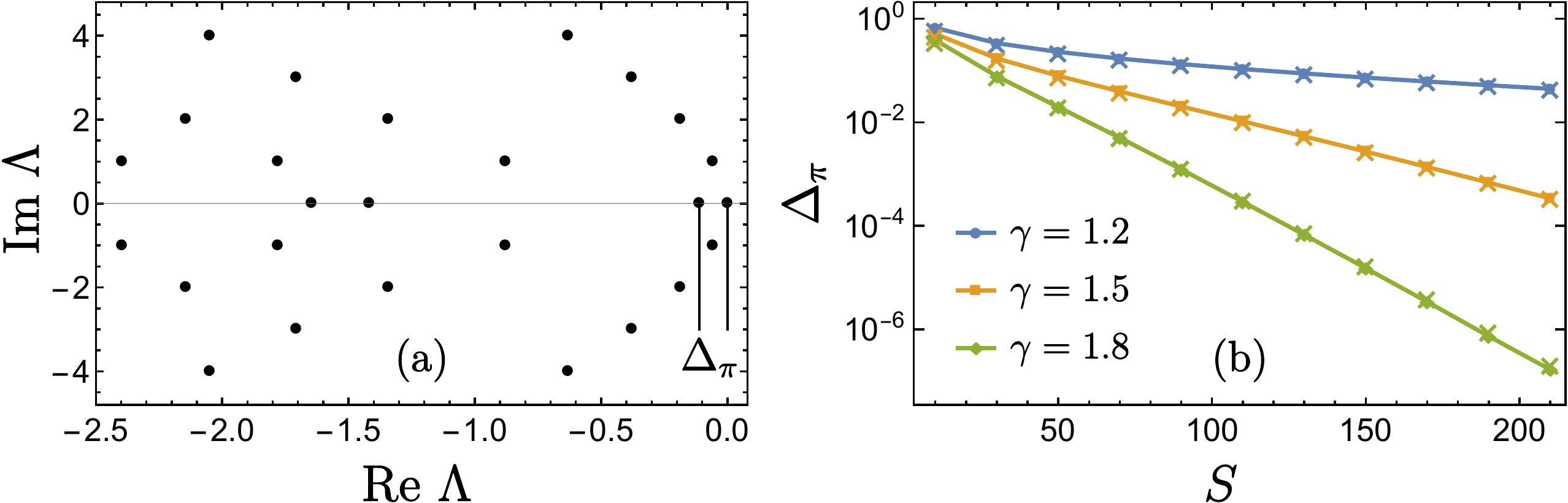}
\caption{\label{fig:southpole_exponential}(a) Low-lying Lindblad spectrum for $S=20$ and $\gamma=2$, showing a small decay rate, $\Delta_{\pi}$, of the eigenstate corresponding to the classical fixed point at the south pole for $\gamma > 1$. (b) Exponential suppression of $\Delta_{\pi}$ with $S$ for different values of $\gamma$. Solid lines are from the Lindblad equation and crosses are from the Fokker-Planck equation [Eq.~\eqref{eq:radial-f}].
}
\end{figure}

The second steady state of the classical spin (at $s_z=-1$) acquires a finite lifetime in the quantum case, as shown in Fig. \ref{fig:southpole_exponential}. The gap $\Delta_{\pi}$ is exponentially small (the lifetime is exponentially large) in $S$.  This decay rate is reproduced by the
Fokker-Planck equation.  However, as $e^{-S}$ is nonperturbative in $1/S$, this splitting cannot be captured by perturbation theory in $1/S$.

\subsection{Hopf bifurcation at $\gamma = 1$}

At the bifurcation point, the classical limit has a marginally stable fixed point at $\theta_c = 0$ and a marginally unstable fixed point at $\tilde{\theta}_c = \pi$. For finite $S$ the steady state in Eq.~\eqref{eq:SmSz_steadystate} is peaked about the former: $f_0(\theta) \sim e^{-r^4/4}$ where $\theta = S^{-1/4} r$. Expressing Eq.~\eqref{eq:radial-f} in terms of $r$ yields the leading-order eigenvalue equation
\begin{equation}
    (\lambda - {\rm i}l) f = 
    \frac{1}{\sqrt{S}} \left[ \frac{{\rm d}^2f}{{\rm d}r^2} + \left( \frac{1}{r} + r^3 \right) \frac{{\rm d}f}{{\rm d}r} - \left( \frac{l^2}{r^2} - 4 r^2 \right) f \right] + O(1/S) \;.
    \label{eq:SmSz_bifpteqn}
\end{equation}
Note that all decay rates scale as $1/\sqrt{S}$, i.e., all the eigenvalues collapse onto the imaginary axis for $S \to \infty$. It is through this spectral collapse that the Lindblad dynamics can reproduce the purely algebraic decay, as detailed in Sec.~\ref{suppsec:algebraic_decay}. The full spectrum includes a second group of eigenvalues obtained by expanding $\theta$ about the south pole, $\theta = \pi - S^{-1/4} r$, which also fall as $1/\sqrt{S}$. The solutions to these eigenvalue equations are related to biconfluent Heun functions \cite{supp_ronveaux1995heun}, but they have to be found numerically.  

\section{Spin model: Animations of wave packet}\label{suppsec:spinmodel_animations}

In the separate “.mp4” files we show simulations of the Fokker-Planck dynamics of the spin model [Eq.~(\ref{eq:SmSz_fokkerplanck})], starting from an initially gaussian probability distribution centered at $\theta = \pi/2$ and $\phi = -\pi/2$, with $S = 50$. For $\gamma = 2$, we observe the wave packet following a classical trajectory to the limit cycle and then slowly diffusing around it, as discussed in Fig.~4 of the main text. For $\gamma = 1$, the center of the wave packet approaches the marginal fixed point at the north pole over a time $t \sim \sqrt{S}$. The angular spread of the steady state is $\Delta \theta \sim S^{-1/2}$ for $\gamma = 2$ and $\Delta \theta \sim S^{-1/4}$ for $\gamma = 1$, as shown in Fig.~3(a) of the main text.



\section{Universal `wedge' spectrum for a fixed point \label{suppsec:fixedpoint_wedge}}

Consider a general case where the classical phase space has a stable fixed point at $\mathbf{z} = \mathbf{z}_c$, and classical trajectories follow the equation of motion ${\rm d}\mathbf{z}/{\rm d}t = \mathbf{u}(\mathbf{z})$, with $\mathbf{u}(\mathbf{z}_c) = 0$. The late-time dynamics close to the fixed point will be described by slow-decaying eigenstates of the quantum Fokker-Planck equation [Eq.~\eqref{eq:fokkerplanck}]. Thus, we expect these eigenstates to be peaked at $\mathbf{z}_c$ and become more pointlike as one approaches the classical limit, i.e., their characteristic width scales as $|| \mathbf{z} - \mathbf{z}_c || \sim 1/\mu$, where $\mu \to \infty$ in the classical limit. (In the spin model $\sqrt{S}$ plays the role of $\mu$.) Substituting $\mathbf{z} := \mathbf{z}_c + \mathbf{r}/\mu$ and $D_{i,j}(\mathbf{z}) := \tilde{D}_{i,j}(\mathbf{z}) / \mu^2$ in Eq.~\eqref{eq:fokkerplanck} one finds that only the linear part of $u(\mathbf{z})$ and the constant part of $\smash{\tilde{D}_{i,j}}(\mathbf{z})$ survive up to $O(1/\mu)$, yielding
\begin{equation}
    \frac{\partial \rho}{\partial t} 
    = \hat{\Lambda} \rho\ 
    := - \sum_i \frac{\partial}{\partial r_i} \Big[ \sum_j J_{i,j} r_j \rho \Big]  + \frac{1}{2} \sum_{i,j} D^c_{i,j} \frac{\partial^2 \rho }{\partial r_i \partial r_j} + O(1/\mu) \;,
\end{equation}
where $J_{i,j} = \left(\partial u_i/\partial z_j\right)|_{\mathbf{z} = \mathbf{z}_c}$ and $D^c_{i,j} = \tilde{D}_{i,j}(\mathbf{z}_c)$. Such a Fokker-Planck operator $\hat{\Lambda}$, with linear drift and constant diffusion, describes an Ornstein-Uhlenbeck process. As shown in Ref.~\cite{supp_leen2016eigenfunctions}, one can construct ladder operators for $\smash{\hat{\Lambda}}$ to obtain the eigenvalues $\Lambda = \sum_i n_i \lambda_i$ with $n_i = 0,1,2 \dots$, where $\lambda_i$ are the eigenvalues of the classical Jacobian $J_{i,j}$.

\section{Parabolic distortion of the limit-cycle branch for the spin model \label{suppsec:limtcyclebranch_perturbthy}}

Here we derive the limit-cycle branch up to $O(1/S)$, Eq.~\eqref{eq:SmSz_lcbranch_pert}, using the semiclassical eigenvalue equation, Eq.~\eqref{eq:radial-f}. This will entail calculating up to second-order correction to the spectrum of a non-Hermitian differential equation. Thus, we first develop perturbation theory for such problems.

Consider a non-Hermitian operator $\hat{A}$ with nondegenerate eigenvalues $\Lambda_i$ and a complete set of right eigenvectors $|\psi_i\rangle$ and left eigenvectors $\langle \chi_i|$, i.e., $\hat{A} |\psi_i\rangle = \Lambda_i |\psi_i\rangle$ and $\langle \chi_i | \hat{A} = \Lambda_i \langle \chi_i|$. The left and right eigenvectors can be chosen bi-orthonormal, i.e., $\langle \chi_i | \psi_j \rangle = \delta_{ij}$. We add a perturbation $\hat{V}$ and look for the new eigenvalues and eigenvectors as a power series,
\begin{equation}
    \big(\hat{A} + \epsilon \hat{V} \big) 
    \sum_{n=0}^{\infty} \epsilon^n \big| \psi_i^{(n)} \big\rangle = 
    \sum_{m=0}^{\infty} \epsilon^m \Lambda_i^{(m)}
    \sum_{n=0}^{\infty} \epsilon^n \big| \psi_i^{(n)} \big\rangle \;,
    \label{eq:perturbation}
\end{equation}
where $\big| \psi_i^{(0)} \big\rangle := |\psi_i\rangle$ and $\Lambda_i^{(0)} := \Lambda_i$. Comparing $O(\epsilon)$ terms we find $(\hat{A} - \Lambda_i) \big| \psi_i^{(1)} \big\rangle = \big[\Lambda_i^{(1)} - \hat{V}\big] |\psi_i\rangle$. Taking inner product with $\langle \chi_i |$ gives the first-order correction to eigenvalues,
\begin{equation}
    \Lambda_i^{(1)} = \langle \chi_i | \hat{V} | \psi_i \rangle \;,
    \label{eq:perturbation_eigval1}
\end{equation}
whereas taking inner product with $\langle \chi_{k \neq i} |$ gives $\big \langle \chi_k \big| \psi_i^{(1)} \big\rangle = \langle \chi_k | \hat{V} | \psi_i \rangle / (\Lambda_i - \Lambda_k)$. Using completeness of the eigenvectors it follows that
\begin{equation}
    \big| \psi_i^{(1)} \big\rangle = 
    \sum_{k \neq i} |\psi_k \rangle \frac{\langle \chi_k | \hat{V} | \psi_i \rangle}{\Lambda_i - \Lambda_k} \;.
    \label{eq:perturbation_eigvec1}
\end{equation}
Next, the $O(\epsilon^2)$ terms in Eq.~\eqref{eq:perturbation} yield $(\hat{A} - \Lambda_i) \big| \psi_i^{(2)} \big\rangle = \big[\Lambda_i^{(1)} - \hat{V}\big] \big| \psi_i^{(1)} \big\rangle + \Lambda_i^{(2)} |\psi_i\rangle$. Taking inner product with $\langle \chi_i |$ and using Eq.~\eqref{eq:perturbation_eigvec1}, we find the second-order eigenvalue correction
\begin{equation}
    \Lambda_i^{(2)} = 
    \big\langle \chi_i \big| \hat{V} \big| \psi_i^{(1)} \big\rangle =  
    \sum_{k \neq i} \frac{\langle \chi_i | \hat{V} | \psi_k \rangle \langle \chi_k | \hat{V} | \psi_i \rangle}{\Lambda_i - \Lambda_k} \;.
    \label{eq:perturbation_eigval2}
\end{equation}

As discussed in Sec. \ref{suppsec:spinmodel_semiclassical_spectrum}, the eigenstates corresponding to the limit cycle are peaked at $\theta_c = \arccos(1/\sqrt{\gamma})$ and have a characteristic width $\Delta\theta = \gamma^{1/4} / \sqrt{S (\gamma-1)}$. Writing $\theta = \theta_c + x \Delta\theta$ in Eq.~\eqref{eq:radial-f} and expanding in powers of $1/S$, we find $(\lambda - {\rm i}l) f = \hat{A} f + (1/\sqrt{S})\hat{V}_1 f + (1/S) \hat{V}_2 f + O(S^{-3/2})$, where
\begin{align}
    \hat{A} &= \big[(\gamma-1)/\sqrt{\gamma} \big] 
    \big( \partial_x^2 + 2 x \partial_x + 2 \big) \;,
    \label{eq:SmSz_lcpert_op0} \\[0.1cm]
    \hat{V}_1 &= -\gamma^{-1/4} \left\{
    (\gamma-1) x \partial_x^2 + \big[\gamma-2 + (\gamma-4) x^2 \big] \partial_x + 2(\gamma-5) x
    \right\} \;,
    \label{eq:SmSz_lcpert_op1} \\[0.1cm]
    \hat{V}_2 &= \frac{\gamma-2}{2} x^2 \partial_x^2 
    + \frac{3(\gamma^2- 5\gamma +3)x + (13-10\gamma)x^3}{3(\gamma-1)} \partial_x 
    - \frac{l^2 (\gamma^2 -4\gamma +5) + 2(13\gamma-17)x^2}{2(\gamma-1)} \;.
    \label{eq:SmSz_lcpert_op2}
\end{align}
The eigenvalues and eigenstates of $\hat{A}$ are given by $\Lambda_j = -2\big[(\gamma-1)/\sqrt{\gamma}] j$, $\chi_j(x) = H_j(x) / (2^j j! \sqrt{\pi})$, and $\psi_j(x) = e^{-x^2} H_j(x)$, $j \in \{0,1,2,\dots\}$, where $H_j$ are the Hermite polynomials. The limit-cycle branch corresponds to $j=0$. Since $\hat{V}_1$ turns an even function to an odd function, and vice versa, it does not alter any of the eigenvalues to $O(1/\sqrt{S})$ from Eq.~\eqref{eq:perturbation_eigval1}. Hence, the lowest-order correction is $O(1/S)$, which comes from first order in $\hat{V}_2$ and second order in $\hat{V}_1$. In particular, for the limit-cycle branch,
\begin{equation}
    \Lambda_0^{(1)} = \langle \chi_0 | \hat{V}_2 | \psi_0 \rangle = \frac{\gamma+2 - l^2 (\gamma^2 - 4\gamma +5)}{2(\gamma-1)} \;.
    \label{eq:SmSz_lcpert_corr1}
\end{equation}
For the correction from $\hat{V}_1$ we use 
\begin{equation}
    \langle \chi_k | \hat{V}_1 | \psi_0 \rangle = 
    -\frac{(\gamma+2) \gamma^{-1/4}}{2^{k-1} k! \sqrt{\pi}} \int_{-\infty}^{\infty} dx \; e^{-x^2} x(x^2-1) H_k(x) = 
    -\frac{\gamma+2}{4 \gamma^{1/4}} \; (2 \delta_{k,1} + \delta_{k,3})
    \label{eq:SmSz_lcpert_V1int}
\end{equation}
in Eq.~\eqref{eq:perturbation_eigval2}, obtaining
\begin{equation}
    \Lambda_0^{(2)} = -\frac{(\gamma+2)\gamma^{1/4}}{4(\gamma-1)} 
    \left[ \langle \chi_0 | \hat{V}_1 | \psi_1 \rangle + 
    \frac{1}{6} \langle \chi_0 | \hat{V}_1 | \psi_3 \rangle 
    \right] = 
    - \frac{\gamma+2}{2(\gamma-1)} \;.
    \label{eq:SmSz_lcpert_corr2}
\end{equation}
Combining Eqs.~\eqref{eq:SmSz_lcpert_corr1} and \eqref{eq:SmSz_lcpert_corr1} we find the limit-cycle branch up to $O(1/S)$,
\begin{equation}
    \lambda_{\text{lc}} \approx {\rm i}l + \frac{\Lambda_0^{(1)} + \Lambda_0^{(2)}}{S} = 
    {\rm i}l - \frac{l^2}{S} \frac{\gamma^2 - 4\gamma + 5}{2(\gamma-1)} \;.
    \label{eq:SmSz_lcbranch_pert}
\end{equation}
Thus, we find that the limit-cycle branch gains a decay rate parabolically dependent on the angular momentum $l$.


\section{Emergence of algebraic decay in the spin model \label{suppsec:algebraic_decay}}

\begin{figure}[!h]
    \centering
    \includegraphics[width=1\columnwidth]{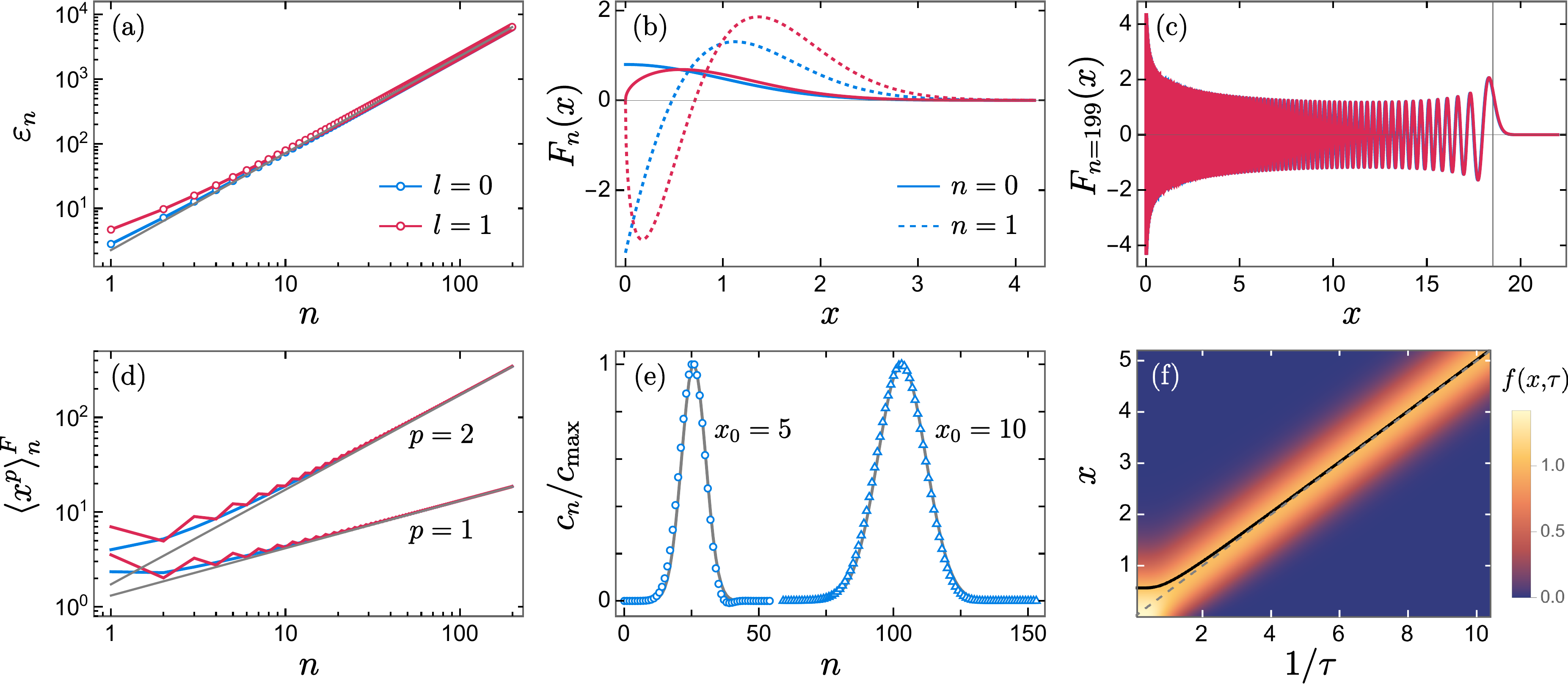}
    \caption{\label{fig:SmSz_algebraic_decay}(a) Rescaled eigenvalues of the leading-order Fokker-Planck equation at $\gamma=1$ [Eq.~\eqref{eq:SmSz_hopfx}]. Gray line shows the power law $\zeta n^{3/2}$ with $\zeta = 2.265$. (b) Low-lying eigenfunctions and (c) Large-$n$ eigenfunction for $l=0$ (blue) and $l=1$ (red). Vertical line in (c) shows the location of the trailing edge at $x = \zeta^{1/3} \sqrt{n}$. (d) First and second moments of the eigenfunctions following the asymptote $(\zeta^{1/3} \sqrt{n})^p$ (gray lines). (e) Expansion coefficients of the localized wavepacket $\smash{f(x) = e^{-2(x-x_0)^2}}$. Gray lines show expected gaussians (see text). (f) The Fokker-Planck equation preserves the width of the wavepacket while its center decays algebraically (solid curve) until reaching the steady state. Dashed line shows the classical algebraic decay, $x = 1/(2\tau)$.
    }
\end{figure}

To understand the algebraic decay to the north pole for $\gamma=1$ in terms of the eigenstates, we write the leading-order eigenvalue equation from Eq.~\eqref{eq:SmSz_bifpteqn} in terms of $x := r^2/2$, obtaining

\begin{equation}
    -\varepsilon f(x) = \mathcal{D} f(x) :=
    x f^{\prime\prime}(x) + \big( 1 + 2x^2 \big) f^{\prime}(x) + \bigg(4x - \frac{l^2}{4x} \bigg) f(x)  
    \;,
    \label{eq:SmSz_hopfx}
\end{equation}
where $\lambda := {\rm i}l -2 \varepsilon / \sqrt{S}$. One can convert Eq.~\eqref{eq:SmSz_hopfx} to a Hermitian eigenvalue equation by a similarity transformation, $f(x) := e^{-x^2/2} F(x)$, which gives 
\begin{equation}
    \varepsilon F(x) = -x F^{\prime\prime}(x) - F^{\prime}(x) + \bigg(x^3 - 2 x + \frac{l^2}{4x} \bigg) F(x) \;.
    \label{eq:SmSz_hopfF}
\end{equation}
One can check that the right-hand side represents a Hermitian operator $\hat{\mathcal{H}}$ by using $\hat{p} := -{\rm i} \partial_x$, which gives $\hat{\mathcal{H}} = \hat{p} \hat{x} \hat{p} + \hat{x}^3 - 2\hat{x} - l^2/(4\hat{x})$. For $l=0$ we have the steady-state ($\varepsilon=0$) solution $F_0(x) \propto e^{-x^2/2}$. Other eigenvalues have to be found numerically, which yields $\varepsilon_n \approx \zeta n^{3/2} [1 + O(1/n)]$ with $\zeta \approx 2.265$ regardless of $l$, as shown in Fig.~\ref{fig:SmSz_algebraic_decay}(a). The eigenfunctions are highly oscillatory for small $x$ and fall off as $e^{-x^2/2}$ beyond $x \approx \zeta^{1/3} \sqrt{n}$ [Figs.~\ref{fig:SmSz_algebraic_decay}(b,c)]. However, remarkably, for $n \gg 1$ they behave like a delta function located at their trailing edge for all moments, i.e., $\int {\rm d}x \; x^p F_n(x) \approx (\zeta^{1/3} \sqrt{n})^p [1 + O(1/n)], \; \forall p \geq 0$ [Fig.~\ref{fig:SmSz_algebraic_decay}(d)], once we have chosen the normalization $\int {\rm d}x F_n(x) = 1$. 

Note that the same normalization does not work for the original eigenstates $f_n(x)$ as $\int {\rm d}x f_n(x) = 0$ whenever $\varepsilon_n > 0$. This follows from the fact that all decaying eigenstates of the Liouvillian are traceless. Generally, how the moments (and expectation of other operators) scale with $n$ depends on the choice of normalization.

As the radial eigenfunctions for different $l$ coincide for large $n$, it suffices to explain the algebraic decay for $l=0$. We consider the initial wavepacket $f(x) = e^{-2(x-x_0)^2}$ with $x_0 \gg 1$, which moves toward $x=0$ while maintaining its shape under the Fokker-Planck dynamics $\partial f / \partial \tau = \mathcal{D} f$, where $\mathcal{D}$ is defined in Eq.~\eqref{eq:SmSz_hopfx} and $\tau := 2 t / \sqrt{S}$. As shown in Fig.~\ref{fig:SmSz_algebraic_decay}(f), the center of the wavepacket falls as $x_* \approx 1/(2 \tau)$ before reaching steady state for $\tau \gtrsim 1$. As $x = \sqrt{S} \theta^2/2$, we recover the classical algebraic decay $1 - s_z \approx \theta^2/2 \approx 1/(4t)$ for $S \to \infty$.

The initial wavepacket can be decomposed in terms of the eigenstates, $F(x) = \sum_n c_n F_n(x)$, where $F(x) = \smash{e^{x^2/2}} f(x)$ is peaked at $x_F = 4x_0/3$ and has a width $\Delta x_F = \smash{\sqrt{2/3}}$. As $F_n(x)$ behaves like a delta function at $x = \zeta^{1/3} \sqrt{n}$, we expect $c_n$ to be peaked at $n_0 = \zeta^{-2/3} x_F^2$ with a spread $\Delta n = 2 \zeta^{-2/3} x_F \Delta x_F$. As shown in Fig.~\ref{fig:SmSz_algebraic_decay}(e), the coefficients are indeed well approximated by such a gaussian for $x_0 \gg 1$. The time-evolved coefficients are given by $c_n(\tau) = c_n(0) e^{-\varepsilon_n \tau} \approx \exp \!\big[\! -(n-n_0)^2/ \Delta n^2 - \zeta n^{3/2} \tau \big]$ up to an overall prefactor. For $\tau \gg \sqrt{n_0} / (\Delta n)^2 = O(1/x_0)$ this profile is peaked at $n_* \approx \zeta^{-2/3} / (2 \tau)^2$, which predicts that the center of the wavepacket decays as $\zeta^{1/3} \sqrt{n_*} \approx 1/(2\tau)$.

\section{Driven-dissipative Bose-Hubbard dimer \label{suppsec:BHdimer}}

\begin{figure}[!h]
\centering
\includegraphics[width=0.35\columnwidth]{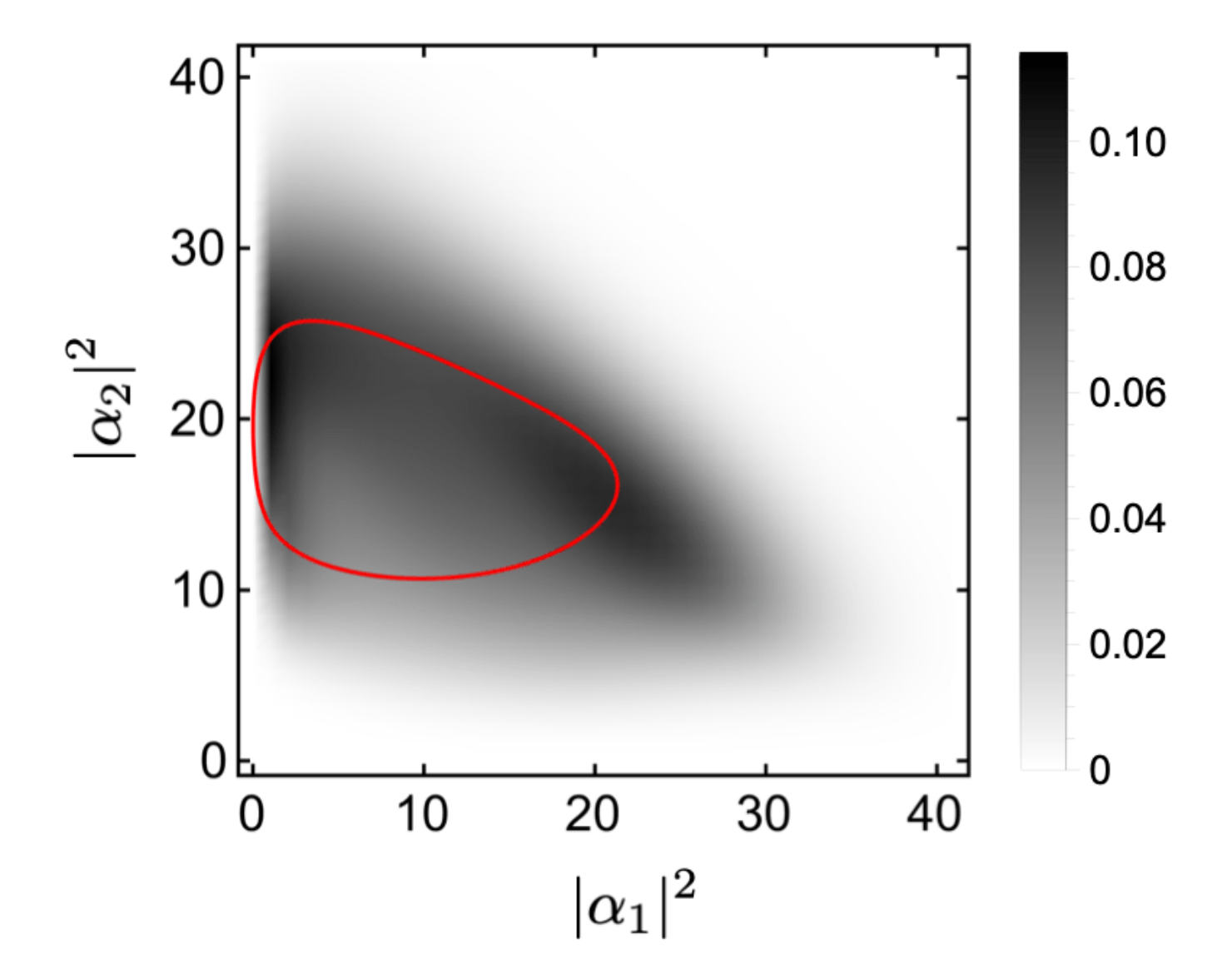}
\caption{\label{fig:BHDimer_limitcycle}Projection of the classical limit cycle (red loop) and distribution of the quantum steady state for $\mu = 5$ (gray cloud). The latter is quantified with the `radial' Wigner function $P(r_1,r_2) := r_1 r_2 \int d\phi_1 \int d \phi_2 W(\alpha_1, \alpha_2)$, where $\alpha_j := r_j e^{i \phi_j}$.
}
\end{figure}

The Bose-Hubbard dimer is described by the Hamiltonian~\cite{supp_giraldo2020driven, supp_giraldo2021chaotic, supp_giraldo2022semiclassical}
\begin{equation}
    \hat{H}_{\text{lab}} = -J (\hat{a}_1^{\dagger} \hat{a}_2 + \hat{a}_2^{\dagger} \hat{a}_1) + 
    \sum_{j=1,2} \left[ 
    \omega \hat{a}_j^{\dagger} \hat{a}_j 
    + \frac{U}{2} \hat{a}_j^{\dagger} \hat{a}_j^{\dagger} \hat{a}_j \hat{a}_j 
    + \big( F_j e^{{\rm i} \omega_d t} \hat{a}_j^{\dagger} + \text{h.c.} \big)
    \right]
    \label{eq:BHDimer_hamil_lab}
\end{equation}
and Lindblad operators $\hat{L}_j = \sqrt{2\kappa}\; \hat{a}_j$, where $J$ is the tunneling parameter, $\omega$ is the natural level spacing, $U$ is the on-site interaction strength, and $\omega_d$ is the drive frequency, $F_j$ are the local drive amplitudes, and $\kappa$ is the loss rate on either site. Going to the rotating frame of the drive by transforming $\hat{a}_j$ to $\hat{a}_j e^{{\rm i} \omega_d t}$, the Hamiltonian reads 
\begin{equation}
    \hat{H} = -J (\hat{a}_1^{\dagger} \hat{a}_2 + \hat{a}_2^{\dagger} \hat{a}_1) + 
    \sum_{j=1,2} \left[ 
    -\Delta \hat{a}_j^{\dagger} \hat{a}_j 
    + \frac{U}{2} \hat{a}_j^{\dagger} \hat{a}_j^{\dagger} \hat{a}_j \hat{a}_j 
    + \big( F_j \hat{a}_j^{\dagger} + \text{h.c.} \big)
    \right] ,
    \label{eq:BHDimer_hamil}
\end{equation}
where $\Delta := \omega - \omega_d$ is the detuning. The Lindblad operators gain a time-dependent phase, which does not affect time evolution. The Heisenberg equations of motion for the bosonic operators are
\begin{equation}
    \frac{d\hat{a}_j}{dt} = 
    -\kappa \hat{a}_j + {\rm i} \big( J \hat{a}_{\bar{j}} + \Delta \hat{a}_j - U \hat{a}_j^{\dagger} \hat{a}_j \hat{a}_j - F_j \big) \;,
    \label{eq:BHDimer_heisenberg}
\end{equation}
for $j=1,2$, where $\bar{1} := 2$ and $\bar{2} := 1$. When the occupation number is large, the evolution corresponds to classical trajectories of $\alpha_j := \langle \hat{a}_j \rangle$ in a four-dimensional phase space where $\langle \hat{a}_j^{\dagger} \hat{a}_j \hat{a}_j \rangle \approx |\alpha_j|^2 \alpha_j$. We have
\begin{equation}
    \frac{1}{\kappa} \frac{d \tilde{\alpha}_j}{dt} = 
    -\tilde{\alpha}_j + {\rm i} \left( 
    \tilde{J} \tilde{\alpha}_{\bar{j}} + \tilde{\Delta} \tilde{\alpha}_j - |\tilde{\alpha}_j|^2 \tilde{\alpha}_j - \tilde{F}_j
    \right) \;,
    \label{eq:BHDimer_classical}
\end{equation}
where $\tilde{J} := J/\kappa$, $\tilde{\Delta} := \Delta/\kappa$, $\tilde{\alpha}_j := \alpha_j \sqrt{U/\kappa}$, and $\tilde{F}_j := F_j \sqrt{U}/\kappa^{3/2}$. 
As the boson occupations grow as $\kappa/U$, we adopt a rescaling parameter $\mu$ and let $U \to U/\mu$ and $F_j \to \sqrt{\mu} F_j$ to explore the approach to the classical limit as $\mu \to \infty$, while keeping Eq.~\eqref{eq:BHDimer_classical} unaltered. 
In Refs.~\cite{supp_giraldo2020driven, supp_giraldo2021chaotic} the classical dynamics were studied in detail for $F_1 = F_2$, for which the system can have two coexisting limit cycles. 
Here, we consider asymmetric drives that give a single limit cycle in certain parameter regimes; see Fig.~(\ref{fig:BHDimer_limitcycle}). Specifically, we have used  $\tilde{J} = -3.5$, $\tilde{\Delta} = 4.5$, $\kappa = 1$,
$\tilde{F}_1= 5$, and $\tilde{F}_2 = 2.5$. 
For finite $\mu$, the Lindblad equation is diagonalized with a cutoff $n_{\text{max}}$ of the occupation number of either site, which is sufficiently large for the results to converge.



The Fokker-Planck equation [Eq.~\eqref{eq:fokkerplanck}] for the phase-space distribution, here the Wigner function $W(\alpha_1, \alpha_2)$ \cite{supp_strunz1998classical}, is
\begin{equation}
    \frac{\partial W}{\partial t} = 
       - \mathbf{u} \cdot \boldsymbol{\nabla} W
    + 4 \kappa W + \frac{\kappa}{2} \nabla^2 W 
   \;,
    \label{eq:Wigner_BHDimer_eom}
\end{equation}
where
\begin{align}
    u_{x_j} &= -\kappa x_j - J p_{\bar{j}} - \Delta p_j + (U/2) \big( x_j^2 + p_j^2 \big) p_j \;,
    \label{eq:Wigner_ux}\\[0.1cm]
    u_{p_j} &= -\kappa p_j + J x_{\bar{j}} + \Delta x_j - (U/2) \big( x_j^2 + p_j^2 \big) x_j - \sqrt{2} F_j \;,
    \label{eq:Wigner_up}
\end{align}
and $\alpha_i := x_i + {\rm i} p_i$.

\begingroup

\endgroup

\end{document}